\documentclass{jpp}
\usepackage{graphicx}
\usepackage{xcolor}
\usepackage{hyperref}

\usepackage[utf8]{inputenc}
\usepackage[T1]{fontenc}
\usepackage{amsmath}

\shorttitle{MHD stability and shaping}
\shortauthor{E. Rodriguez}

\title{MHD stability and the effects of shaping: a near-axis view for tokamaks and quasisymmetric stellarators}

\author{Eduardo Rodr\'{i}guez\aff{1}
  \corresp{\email{eduardor@princeton.edu}}}

\affiliation{\aff{1} IPP Max Planck Institute for Plasma Physics, Greifswald, Germany
}

\begin{document}

\maketitle

\begin{abstract}
How much does the cross-section of a toroidal magnetic field configuration tell us about its MHD stability? It is generally believed that positive triangularity (typically leading to bean-shaped cross-sections with their indentation on the inboard side in stellarators) contributes positively to MHD stability. In this paper, we explore the basis of this statement within a near-axis description for axisymmetric and quasisymmetric magnetic configurations. In agreement with the existing literature, we show that positive triangularity stabilises vertically elongated tokamaks. In quasisymmetric stellarators, the toroidal asymmetry of flux surfaces modifies this relation. The behaviour of stellarator-symmetric, quasisymmetric stellarators can still be described in terms of the shape of one of their up-down symmetric cross-sections. However, we show that for a sample of quasisymmetric configurations, the positive-bean-shaped cross-sections do not contribute positively to stability. Unlike in the axisymmetric case, we also learn that finite $\beta$ can improve stability even without magnetic shear. 
\end{abstract}

\section{Introduction}
The quest for controlled thermonuclear fusion has seen renewed interest over the last decade. This has led to a revival of concepts other than the tokamak. The latter relies on toroidal magnetic fields with a spatial toroidal symmetry to confine a hot plasma within \citep{mukhovatov1971,wessonTok}. It is useful to consider devices in which this symmetry requirement is relaxed, i.e., stellarators \citep{spitzer1958,boozer1998,helander2014}. { The three-dimensional nature of stellarators provides them with a freedom necessary to avoid many limiting features of axisymmetric magnetic fields.} Most importantly, large currents are no longer needed to hold the plasma, thus minimising the possibility of violent disruptions \citep{schuller1995}. 
\par
Finding attractive forms of stellarators that serve as magnetic confinement devices requires a dedicated effort. At the most basic level, the fields must be capable of confining collisionless charged particles for long enough \citep{mynick2006}. This requirement significantly restricts the space of stellarators, singling out a particular class of fields labelled \textit{omnigeneous} \citep{bernardin1986,cary1997,hall1975,landreman2012,helander2014}. In this paper, we consider a particular group of omnigeneous stellarators, a most immediate generalisation of axisymmetry, known as \textit{quasisymmetric stellarators} \citep{boozer1983,nuhren1988,rodriguez2020,burby2020}. The defining property of this class is a particular form of hidden symmetry, which makes the magnitude of the magnetic field, $|\mathbf{B}|$, symmetric, but not necessarily $\mathbf{B}$. By Noether's theorem, such partial symmetry is sufficient to prevent rapid loss of particles in the small-gyroradius limit. 
\par
Confinement of single particles is not the only desired property of the stellarator: plasma stability, coil complexity, turbulent transport, etcetera, are also aspects of importance. This list of properties, alongside advances in computation, have naturally led to \textit{optimisation} as the primary approach to the design of stellarators. This approach has proven to yield practical results \citep{beidler1990,anderson1995,Garabedian2008,Zarnstorff2001,Najmabadi2008}. However, it is not entirely satisfactory. The complexity of optimisation is prohibitive when attempting to comprehend the origin of the obtained designs \citep{rodriguez2021opt}. This is partly a result of the space of optimisation having many local minima \cite{bader2019,henneberg2021}. Some fundamental insight is necessary to interpret results and guide optimisation in such a space.
\par
Understanding the relation between the properties imposed on stellarators is crucial. If two properties require similar or opposite magnetic field features, we should know it and perform optimisation accordingly. Intuition on these property trade-offs has developed over years of optimisation efforts. It is, for instance, believed that bean shapes (i.e., sizeable positive triangularity) favour MHD stability. { This general wisdom accrued over years of stellarator optimisation \citep{nuhrenberg2010} which regularly found these features, as well as more dedicated works \citep{lortz1976,nuhrenberg1986}.} We put this observation to the test in this paper for axisymmetric and quasisymmetric stellarators.
\par
To explore the question, we use two main ingredients. First, as a measure of MHD stability, we employ the Mercier criterion. We present the basics of this measure in Section 2, together with the main theoretical framework for the paper: the near-axis expansion { in inverse coordinates}. The latter provides a simplifying description of the geometry and governing equations asymptotically in the distance from the centre of the stellarator. This framework becomes an ideal basis for analysing the problem. Section 3 considers the case of axisymmetry, which is often taken as reference to develop intuition on property trade-offs. We show that there is indeed a positive link between positive triangularity and stability in this scenario, reproducing the results of existing literature. That enables us to treat the case of quasisymmetric stellarators in Section 4 analogously. Conventional shaping-stability intuition generally fails there, with negative triangularity being stabilising in practical examples. We close with some concluding remarks. 

\section{Mercier criterion and near-axis framework}
For the purpose of this paper we shall consider static equilibria with isotropic pressure \citep{wessonTok,freidberg2014}, in which magnetic field lines live on nested toroidal magnetic flux surfaces \citep{grad1967b,helander2014}, labelled by the variable $\psi$ ($1/2\pi$ the toroidal magnetic flux). Given such an equilibrium, we are interested in knowing whether it is MHD-unstable or not. Although stellarators have proven certain non-linear resilience to instability, an unstable configuration still represents a physically unattainable configuration. Knowing this is important at least to understand if the stellarator properties carefully designed will reliably hold or not. 
\par 
There is not a single way to study MHD stability. In this paper we turn to two scalar criteria for instability: the magnetic well, $V''>0$ \citep{greene1997}, and the Mercier criterion,  $D_\mathrm{Merc}<0$, \citep{mercier1962,mercier1974,greene1962,bauer2012,freidberg2014}. The latter is a sufficient condition for the occurrence of an \textit{interchange} instability; namely, an instability that displaces the plasma without significantly bending field-lines. A configuration is necessarily unstable if $D_\mathrm{Merc}<0$ (but $D_\mathrm{Merc}>0$ does not guarantee stability, as it does not negate ballooning instability \citep{correa1978,freidberg2014,dewar1983,connor1978,nuhrenberg1988}). The scalar $D_\mathrm{Merc}$ involves multiple integrals over the fields and geometry of the configuration and can be found in the literature \citep{bauer2012,greene1962,correa1978,nuhrenberg1988,zocco2018} (we include it in Appendix~\ref{sec:appMercier} for completeness). The magnetic well criterion \citep{greene1997} is nothing but the small plasma-$\beta$, zero magnetic shear limit of the Mercier criterion. The complexity of both expressions in these stability criteria naturally require a simplified framework in which to understand its underlying structure and relation to field properties.
\par
We adopt for that purpose the so-called near-axis framework, pioneered by \cite{mercier1962} and \cite{Solovev1970}, in which the stellarator is considered asymptotically near its magnetic axis, where the stellarator is particularly simple. This way of approaching the problem has seen a recent revival for theoretical and practical stellarator design \citep{garrenboozer1991a,landreman2019,jorge2020a,landreman2022,rodriguez2022phase}. In its original form, often referred to as the \textit{direct-coordinate} approach, one expands all the equations governing the magnetic field in the distance from the axis involving the shape of flux surfaces directly. That way, shaping and stability were related, through the Mercier criterion, by the works of \cite{Solovev1970,lortz1976,mikhailovskii1979,shafranov1983,freidberg2014}, and most recently \cite{jorge2020b,kim2021}. The emphasis on the geometry of the stellarator in this asymptotic description does however not lend itself straightforwardly to describing stability and guiding centre dynamics. The latter naturally involve $|\mathbf{B}|$, which is a quantity not readily accessible in this framework. In particular, this complication makes the description of optimised stellarators, such as omnigeneous \cite{bernardin1986,cary1997,hall1975,landreman2012} or quasisymmetric ones \citep{boozer1983,nuhren1988,rodriguez2020}, challenging.
\par
To bypass these limitations and describe axisymmetric and quasisymmetric stellarators in this paper, we adopt the \textit{inverse{ -coordinate} near-axis expansion} \cite{garrenboozer1991a,landreman2018a,landreman2019}. Its main difference to the direct{-coordinate} approach is that it involves $|\mathbf{B}|$ explicitly in the problem, {rather than the geometry of flux surfaces}. It does so by treating Boozer coordinates \citep{boozer1981} $\{\psi,\theta,\phi\}$ as an independent set, enabling the use of $\epsilon=\sqrt{2\psi/\bar{B}_0}$ (a pseudo-radial coordinate) as an expansion parameter. Here $\bar{B}_0$ is the average magnetic field magnitude along the magnetic axis and $\psi=0$ on it. In terms of this $\epsilon$, in the near-axis expansino all fields are expanded in power series of the form,
\begin{equation}
f(\psi,\theta,\phi)=\sum_{n=0}^\infty\epsilon^n\sum_m^n{}'\left[f_{nm}^c(\phi)\cos m\theta+f_{nm}^s(\phi)\sin m\theta\right],
    \label{eqn:fExpansion}
\end{equation}
where the sum $\sum'$ is over even or odd positive numbers (including 0) up to $n$ depending on the parity of $n$ \citep{garrenboozer1991a}. As the paper focuses on axisymmetric and quasisymmetric configurations, the magnetic field magnitude has expansion parameters independent of the toroidal angle,
\begin{equation}
|\mathbf{B}|=B_0(1+\epsilon\eta\cos\chi)+\epsilon^2(B_{20}+B_{22}^C\cos2\chi+B_{22}^S\sin2\chi)+\dots. \label{eqn:Bnae}
\end{equation}
Here the helical angle $\chi=\theta-N\phi$ takes the place of $\theta$, where $N\in\mathbb{N}$ is the self-linking number of the magnetic axis \cite{rodriguez2022phase}. This form allows us to include quasi-helical configurations in the formalism (quasisymmetric stellarators in which the contours of $|\mathbf{B}|$ close helically for $N\neq0$). 
\par
Using the inverse-coordinate approach relegates the shaping of magnetic flux surfaces to a secondary role. Flux surfaces are nevertheless described around the magnetic axis, using the Frenet-Serret vectors of the axis (see \cite{landreman2019}) as a basis, as \begin{equation}
    \mathbf{x}-\mathbf{r}_0=X(\psi,\chi,\phi)\hat{\kappa}+Y(\psi,\chi,\phi)\hat{\tau}+Z(\psi,\chi,\phi)\hat{\mathbf{b}}, \label{eqn:surfaceNAE}
\end{equation}
for all $\chi$ and $\phi$ (at constant $\psi$), where $\mathbf{r}_0$ represents the magnetic axis, $\hat{\kappa}$ is the normal to the curve, $\hat{\tau}$ the binormal, and $\hat{\mathbf{b}}$ its tangent. In any plane normal to the axis (and disregarding the function $Z$), $X$ and $Y$ describe the shape of the cross-sections. Our task in this paper is then to interpret the shaping through $X,~Y$ and $Z$, and relate it to the Mercier criterion.  
\par
Details of the near-axis expansion in the so-called inverse{-coordinate} approach may be found in the original paper by \cite{garrenboozer1991a}, or later applications and extensions \cite{landreman2018a,landreman2019,rodriguez2020i}. Here we shall not rederive these results but rather apply them; the reader may refer to these if necessary. The magnetic well on axis \citep{landreman2020} is,
\begin{equation}
    V''=\frac{8\pi^2 G_0}{B_0^3}\left[\frac{3\eta^2}{2}-\frac{2B_{20}}{B_0}-\frac{p_2}{B_0^2}\right], \label{eqn:magWellNAE}
\end{equation}
where the pressure gradient (to leading order) is given by $p_2$ (from $p=p_0+\epsilon^2 p_2+\dots$, which includes for simplicity a factor of $\mu_0$), and $G_0$ is the poloidal current linked to the torus. As we have hinted, the expression is simple and involves $|\mathbf{B}|$ quite directly. Here the parameter $B_{20}$ may be interpreted as a measure the depth of the magnetic well \cite{freidberg2014}. 
\par
The pressure gradient in Eq.~(\ref{eqn:magWellNAE}) is somewhat deceptive. We remind ourselves that using $V''$ as a criterion is sensible only in the vacuum limit of the \textit{Mercier criterion}. Thus, for a finite-$\beta$ equilibrium, the Mercier criterion reads, following \cite{landreman2020} and taking $B_0=1$ for simplicity,
\begin{equation}
    D_\mathrm{Merc}=\frac{G_0^2p_2}{\pi^2\epsilon^2}\left\{\frac{3\eta^2}{2}-2B_{20}-2p_2\left(1+\frac{2G_0^2\eta^2}{\Bar{\iota}_0^2}\mathcal{I}\left[\frac{\eta}{\kappa},\sigma\right]\right)\right\}, \label{eqn:mercierCriterion}
\end{equation}
where,
\begin{equation}
    \mathcal{I}\left[\Bar{\eta}=\frac{\eta}{\kappa},\sigma\right]=\frac{1}{2\pi}\int_0^{2\pi}\frac{\Bar{\eta}^4+\Bar{\eta}^2+\sigma^2}{\Bar{\eta}^4+2\Bar{\eta}^2+(1+\sigma^2)}\mathrm{d}\phi, \label{eqn:DmIint}
\end{equation}
$\kappa$ is the curvature of the magnetic axis (a function of the toroidal angle $\phi$), $\sigma=Y_{11}^c/Y_{11}^s$ (see the notation of Eq.~(\ref{eqn:fExpansion})) is related to the shaping of flux surfaces, and $\Bar{\iota}_0$ is the rotational transform on-axis (in the quasi-helical case $\iota_0-N$). The structure of the leading order Mercier criterion is, up to the dependence on the pressure gradient, of the same form as the magnetic well, depending critically on $B_{20}$.

\section{Case of axisymmetry}\label{sec:tokamak}
In the previous section we presented the Mercier criterion in its near-axis form following \cite{landreman2020}. Our goal now is to interpret Eq.~(\ref{eqn:mercierCriterion}) meaningfully in terms of the shaping of cross-sections. To do so, we must consider two essential points. First, we must know which choice of parameters within the near-axis description serve as a minimal, consistent parametrisation of our near-axis equilibrium. In this section, we focus on these choices for axisymmetry, leaving quasisymmetry for the next section. Secondly, given such a set, we need to connect them within the near-axis expansion to the most common notions of cross-section shapes such as triangularity. Once this has been achieved, we will be in a position to discuss MHD stability and its relation to the shape of the plasma cross-sections. 

\subsection{Parametrisation of configurations} \label{sec:paramsNAE}
Let us describe axisymmetric configurations uniquely within the near-axis framework. We present the most conventional choice of parameters, which we must however point is not unique. 
\par
The shape of the magnetic axis is the primary ingredient in the expansion, but in the case of axisymmetry it must be a circle. We normalise its radius to $R_0=1$, as we do with the magnetic field on axis, $B_0=1$, leading by Amp\`{e}re's la to a poloidal current $G_0=1$. At first order, the parameters $\eta$ (leading-order $|\mathbf{B}|$ mirror ratio, see Eq.~(\ref{eqn:Bnae})) and $\sigma$ (measure of the up-down asymmetry) describe rather explicitly elliptical flux surfaces (as we will later see). The choice of the toroidal current density on axis, $I_2$, then provides a finite rotational transform on axis. 
\par
Finally, at second order, four parameters are necessary: the pressure gradient, $p_2$, and the second order harmonics of $|\mathbf{B}|$, $B_{20}$, $B_{22}^C$ and $B_{22}^S$. Not all choices of these natural parameters constitute valid equilibria, though. The force balance condition imposes a linear constraint, Eq.~(\ref{eqn:B20eqApp}), making  only three of them truly independent. Different choices of independent parameters are suitable for studying different equilibrium properties. In our case, we must find the combination of natural parameters $p_2$ and $|\mathbf{B}|$ harmonics that directly relates to the shaping of cross-sections, so that we may use them as the independent set. That is the next task. One may ask why the direct-coordinate near-axis approach was chosen if the geometry was wanted explicitly. The answer is that we want to have the capacity to constrain $|\mathbf{B}|$ directly and simply, and within these constraints, see how the geometry arises.


\subsection{Shapes within the inverse{-coordinate} near-axis framework}
We must thus construct geometric notions for the cross-sections within the near-axis framework. These include concepts such as ellipticity, triangularity and up-down symmetry breaking. Although the description in this section is concerned with axisymmetry, the description of shaping presented generalises straightforwardly to the quasisymmetric case.  
\par

\subsubsection{{ First-order shaping}: ellipticity} \label{sec:ellip}
As is well-known, near the magnetic axis, flux surfaces have elliptic cross-sections. In the near-axis expansion these are described at first order in the expansion in terms of the parameters $\eta$ and $\sigma$. The shapes in the plane normal to the axis (in our tokamak, slices at constant cylindrical angle) are described as
\begin{equation}
    \mathbf{x}=-\epsilon\eta\cos\chi\hat{R}+\frac{\epsilon}{\eta}(\sin\chi+\sigma\cos\chi)\hat{z}, \label{eqn:ellipEq}
\end{equation}
where $\hat{R}$ is the unit vector in the major radius direction, $\hat{z}$ the unit vector in the vertical cylindrical direction, $X_{11}^c=\eta$ and $Y_{11}^s=1/\eta$ \citep{landreman2019}. 
\par
\begin{figure}
    \centering
    \includegraphics[width=0.2\textwidth]{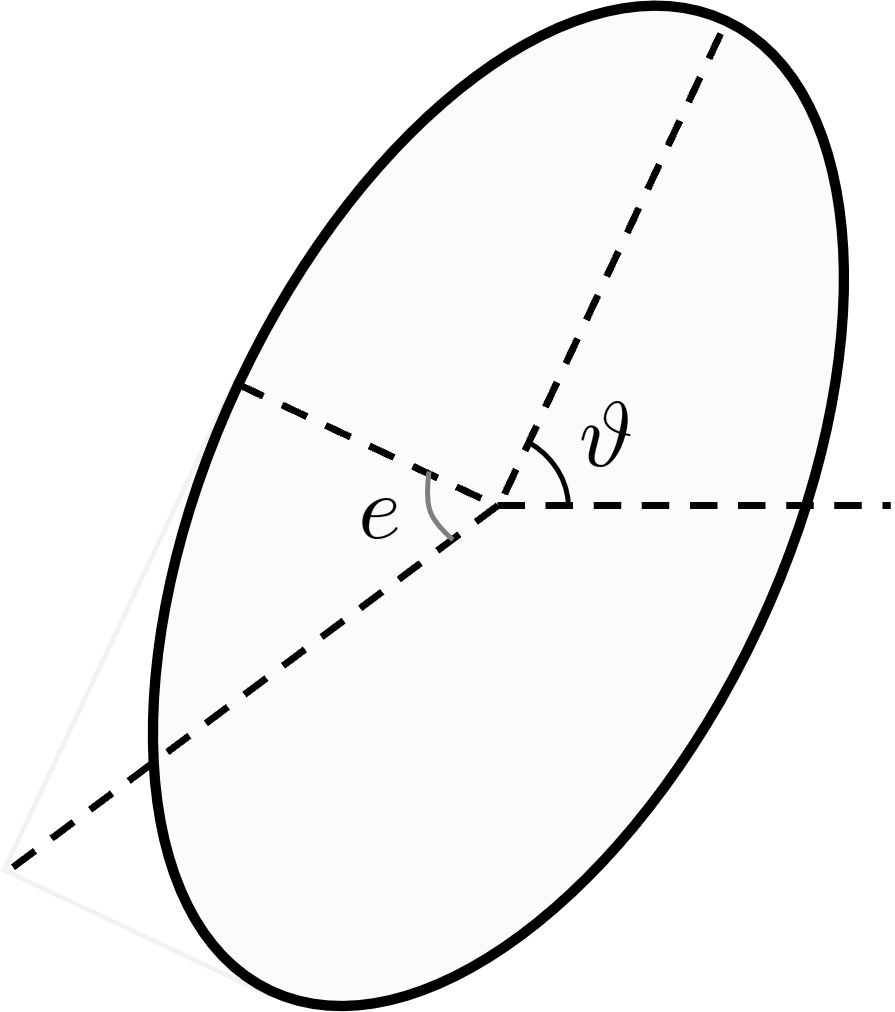}
    \caption{\textbf{Elliptical shapes and angles.} Diagram showing an ellipse framed in the normal Frenet-Serret frame where the ellipse rotation angle $\vartheta$ and elongation angle $e$ are defined. These two angles uniquely characterise ellipses (up to a scale). }
    \label{fig:ellipseAng}
\end{figure}
In terms of the elongation (defined as the ratio of the major to the minor radius of the ellipse, $\mathcal{E}$) and the rotation angle $\vartheta$ with respect to $\hat{\kappa}$, and defining the angle $\mathcal{E}=\tan e$ (see Fig.~\ref{fig:ellipseAng}),
\begin{subequations}
    \begin{gather}
        \sin 2e=\frac{2\eta^2}{1+\sigma^2+\eta^4}, \label{eqn:ellipAng1}\\
        \tan 2\vartheta=\frac{2\sigma\eta^2}{\eta^4-1-\sigma^2}. \label{eqn:ellipAng2}
    \end{gather}\label{eqn:ellipAng}
\end{subequations}
For the details on how one arrives at Eqs.~(\ref{eqn:ellipAng}), see Appendix~\ref{sec:appEllip}. From the above, it is clear that $\sigma$ rotates the ellipse respect to the $(\hat{\kappa},\hat{\tau})$ Frenet-Serret frame, in the small $\sigma$ limit linearly, and thus is a measure of up-down asymmetry. However, it also affects elongation through the denominator in Eq.~(\ref{eqn:ellipAng1}). Only in the limit of $\sigma=0$, for which $\vartheta=0,~\pi/2$, $\mathcal{E}=\eta^2,~1/\eta^2$ respectively, and elongation just depends on $\eta$. This limit allows us to interpret $\eta$ as a measure (approximate) of elongation, rigorously true in the up-down symmetric limit. 
\par
In an up-down asymmetric scenario, the distortion of $\eta$ as a measure of elongation make $\vartheta$ and $e$ become the natural shaping parameters. Expressing the near-axis expansion in terms of these parameters yields, however, highly complicated and non-linear expressions that necessarily require of numerical tools to handle. Thus, we shall use $\eta$ and $\sigma$ as our parameters, capitalising on their approximate geometric meaning for interpretation. 
\par

\subsubsection{{ Second-order shaping}: triangularity}
\begin{figure}
    \centering
    \includegraphics[width=0.4\textwidth]{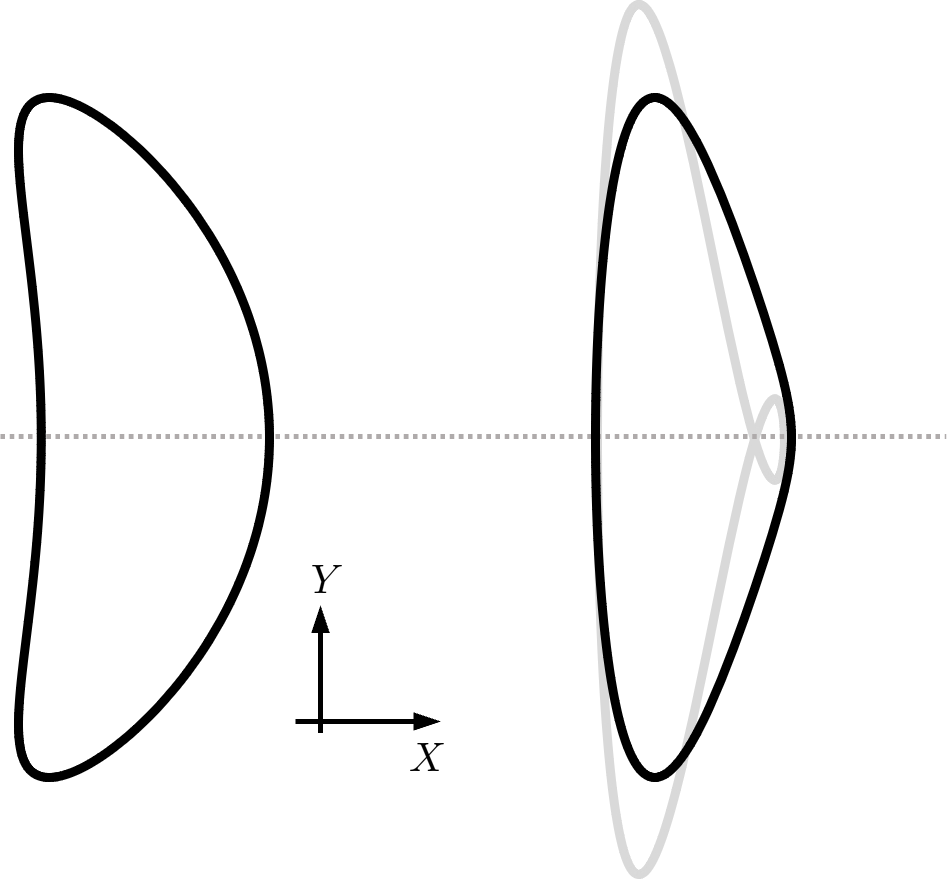}
    \caption{\textbf{Examples of triangular cross-sections.} The diagram shows two examples of triangular cross-sections (in black), {constructed with second-order shaping with $X_{22}^C<0$ (left) and $Y_{22}^S>0$ (right), for the same underlying elliptical shape}. The stellarator literature refers to these shapes as \textit{bean} and \textit{D} shapes (often clearer for less elongated cross-sections), respectively. The grey contour shows how excessive shaping can lead to a pathological cross-section in which the surface self-intersects.}
    \label{fig:triangDiagShape}
\end{figure}

In increasing order of complexity, the next family of shapes after ellipses is  \textit{triangularity}. Figure~\ref{fig:triangDiagShape} shows cross-sections with non-zero triangularity, formally arising at second order in the near-axis expansion through $X_2=X_{20}+X_{22}^C\cos2\chi+X_{22}^S\sin2\chi$, and equivalently for $Y_2$. 
Before constructing a quantitative measure of triangularity, let us first familiarise ourselves with each of these shaping coefficient at second order considering a frame aligned ellipse at first order. 
\par
Take first the $X_{22}^C\cos2\chi$ term (see the leftmost cross-section in Fig.~\ref{fig:triangDiagShape}). The magnitude of $X_{22}^C$ gives the `bean-shape' of the cross-section, becoming ever `more triangular' as its magnitude is increased, eventually developing a characteristic dimple or indentation. Geometrically, the indentation of the bean shape appears (see Fig.~\ref{fig:triangDiagShape}) upon crossing the threshold $\epsilon X_{22}^C/X_{11}^C\geq1/4$.\footnote{This condition follows from assessing the existence of turning points in $X$, $\partial_\chi(X_{11}^C\cos\chi+\epsilon X_{22}^C\cos2\chi)\stackrel{!}{=}0$, so that $\cos\chi=-X_{11}^C/4\epsilon X_{22}^C$ will only have real solutions for $\epsilon X_{22}^C/X_{11}^C\geq1/4$. { Note that the dimple may appear either on the inboard or outboard side, depending on the sign of $X_{22}^c$ and, thus, the sense of the bean shape.}} The strength of the shaping is thus measured by $\epsilon X_{22}^C/X_{11}^C$, and increases away from the axis. 
\par
The meaning of $Y_{22}^S$ is not dissimilar and is also related to what is commonly perceived as triangularity. However, it manifests as a more \textit{D} looking shape (see the rightmost plot in Fig.~\ref{fig:triangDiagShape}). As $\epsilon Y_{22}^S$ becomes larger, the shape becomes more and more triangular until it reaches a critical value beyond which the cross-section self intersects (see Fig.~\ref{fig:triangDiagShape}). The critical point, i.e., the first instance in which the cross-section crosses the $Y=0$ line thrice, corresponds to $\epsilon Y_{22}^S=Y_{11}^S/2$. The strength of the shaping is then $\epsilon Y_{22}^S/Y_{11}^S$, which limits the near-axis description to $\epsilon<\epsilon_\mathrm{max}=Y_{11}^S/2Y_{22}^S$. A similar limit exists for $X_{22}^c$ to prevent the indentation of the bean shape from being too large so that nested surfaces touch and eventually cross each other. Both of these are a geometric interpretation of the measure $r_c$ introduced by \cite{landreman2021a}. 
\par
Although in different flavours, both of these components bring triangularity in. So far, we have been vague on what we mean by triangularity, and we must invoke a more rigorous definition for a quantitative consideration. We define triangularity as the relative displacement of the vertical tips of the cross-section from the cross-section mid-point divided by the width of the cross-section. {It is a measure of left-right asymmetry of the cross-sections.} We choose a positive value to indicate a relative displacement in the direction of $\hat{\kappa}$ (in the tokamak $-\hat{R}$). Following this definition (calculation details may be found in Appendix~\ref{sec:appTriang}), the asymptotic form of triangularity, $\delta_\mathrm{tok}$, may be written as 
\begin{equation}
    \delta_\mathrm{tok}\approx 2\epsilon\left(\frac{Y_{22}^S}{Y_{11}^S}-\frac{X_{22}^C}{X_{11}^C}\right), \label{eqn:triangAsymp}
\end{equation}
which involves the shaping strength fractions previously obtained, with the negative sign being consistent with the picture in Fig.~\ref{fig:triangDiagShape}.
\par
{The same way that this notion of triangularity, $\delta_\mathrm{tok}$, measures the degree of left-right asymmetry, we define another geometric parameter that we call vertical triangularity, $\delta_y$, which measures the degree of up-down asymmetry.} In this case, in terms of the shaping parameters $X_{22}^s$ and $Y_{22}^c$,
\begin{equation}
    \delta_y\approx2\epsilon\left(\frac{Y_{22}^C}{Y_{11}^S}+\frac{X_{22}^S}{X_{11}^C}\right). \label{eqn:triangYAsymp}
\end{equation}
Note the sign difference with Eq.~(\ref{eqn:triangAsymp}) (in fact, one can see a similar sign in \cite[Eqs.~(5.1)-(5.2)]{rhodes2017}). We have taken the convention that a positive $\delta_y$ indicates an upwards bulging (meaning in the direction of the binormal to the axis). 
\par
When the underlying elliptic shape is not frame-aligned, the description of the shaping becomes significantly more complex, as the components of $X_2$ and $Y_2$ mix together. Appendix~\ref{sec:appTriang} briefly describes how to deal with this situation by defining effective  measures $\delta_\mathrm{tok}$ and $\delta_y$ in the rotated frame. We will give a numerical example of what this means later.

\subsubsection{{ Second-order shaping}: Shafranov shift}
So far, we have said little regarding the relative position of flux surfaces, which the \textit{Shafranov shift} \citep{shafranov1963,wessonTok} is a measure of. Classically, this is defined as the relative shift of the centres of circular cross-sections in the large-aspect-ratio-limit of an axisymmetric configuration. In a more general stellarator, the shift becomes ambiguous, as the centre of generally shaped cross-sections (other than ellipses and circles) is non-unique. We opt to define it as,
\begin{subequations}
    \begin{gather}
        \Delta_x=X_{20}+X_{22}^C, \label{eqn:shafShiftNAEX}\\
        \Delta_y=Y_{20}+Y_{22}^C,\label{eqn:shafShiftNAEY}
    \end{gather}\label{eqn:shafShiftNAE}
\end{subequations}
where the near-axis expansion parameters have been directly used. In an up-down symmetric configuration, $\Delta_x$ describes the displacement of the cross-section midpoint along the up-down symmetry line from one surface to the next. The vertical portion, $\Delta_y$, has a similar interpretation along the binormal ($Y$). Appendix~\ref{sec:appShafShift} motivates the form of this definition of Shafranov shift beyond its simple geometric interpretation, following the work in \cite{rodriguez2022}. This form of the Shafranov shift reduces to the correct tokamak definition (see \cite{landreman2021a}). As we shall be primarily concerned with the up-down symmetric form of the problem, we shall refer to $\Delta_x$ as the Shafranov shift. 

\subsection{MHD stability and cross-section shapes}
The geometry parameters in the previous subsection constitute the appropriate shaping-related parameters in terms of which we ought to express the near-axis expansion. For simplicity, at lowest order, we choose to use parameters $\eta$ and $\sigma$ explicitly. { We shall, unless otherwise stated, focus on up-down symmetric configurations, and so take $\sigma=0$ and $\delta_y=0$.} Having the magnitude of the pressure gradient, $p_2$, explicitly involved is often convenient, as it allows us to study the effect of other parameters on a configuration that needs to support a prescribed `pressure profile'. This leaves one of the two geometric measures, either the Shafranov shift, $\Delta_x$, or the triangularity, $\delta=\delta_\mathrm{tok}/\epsilon$, to complete the parametrisation of the near-axis configuration. Choosing one explicitly will make the other adjust self-consistently in a concealed way so that it  complies with equilibrium. 
\par
To express everything in terms of these parameters, we must relate them to $|\mathbf{B}|$ components through $\{X_2,Y_2\}$ coefficients. These relations form part of the near-axis expansion, and are given explicitly in Appendix~\ref{sec:equationsNAE}. Although algebraically involved, computational algebra may handle them straightforwardly. { That way, we first present three equivalent forms of $B_{20}$ in terms of different relevant second-order parameters, each with a different physical interpretation,}
\begin{subequations}
\begin{align}
    B_{20}=&-\left[1+\frac{(1+\eta^4)^2}{(3+\eta^4)I_2^2}\right]p_2+\frac{3\eta}{2}\frac{1-\eta^4}{3+\eta^4}\delta+\eta^2\left(\frac{4+\eta^4}{3+\eta^4}-\frac{I_2^2}{1+\eta^4}\right),\label{eqn:B20pd}\\
    =&-\left[1+\frac{(1+\eta^4)^2}{4\eta^4I_2^2}\right]p_2+\frac{3}{2}\frac{\eta^4-1}{\eta^4}\Delta_x+\eta^2\left(\frac{1+4\eta^4}{4\eta^4}-\frac{I_2^2}{1+\eta^4}\right),\label{eqn:B20pD}\\
    =&2\Delta_x\left[1+\frac{(3+\eta^4)I_2^2}{(1+\eta^4)^2}\right]+\frac{\eta}{2}\left[1+\frac{4\eta^4I_2^2}{(1+\eta^4)^2}\right]\delta+\eta^2\left(1-\frac{(2+\eta^4)I_2^2}{(1+\eta^4)^2}\right). \label{eqn:B20Dd}
\end{align}
\end{subequations}
Following this, the Mercier criterion, Eq.~(\ref{eqn:mercierCriterion}), can be written as,
\begin{subequations}
\begin{align}
    \frac{\epsilon^2\pi^2 D_\mathrm{Merc}}{|p_2|G_0^2}=&\frac{(\eta^2-1)^2(1+\eta^4)^2}{(1+\eta^2)(3+\eta^4)I_2^2}p_2+3\eta\delta\frac{1-\eta^4}{3+\eta^4}+\frac{\eta^2}{2}\left(\frac{7+\eta^4}{3+\eta^4}-\frac{4I_2^2}{1+\eta^4}\right), \label{eqn:axisDMercpd}\\
    =&\frac{(\eta^2-1)(2\eta^2+1)(1+\eta^4)^2}{2\eta^4(1+\eta^2)I_2^2}p_2+3\Delta_x\frac{\eta^4-1}{\eta^4}+\frac{1}{2\eta^2}\left(1+\eta^4-\frac{4\eta^4I_2^2}{1+\eta^4}\right).\label{eqn:axisDMercpD}
\end{align}
\end{subequations}
The effects of shaping are buried in each of the terms of these expressions, especially their sign. If the factor multiplying a particular parameter in the Mercier criterion is positive, then the geometric or physical property represented by the parameter can be said to have a stabilising effect. 
\par
Consider first Eq.~(\ref{eqn:B20Dd}), which describes the direct effect of cross-section shaping on $B_{20}$. Because the factors multiplying both $\delta$ and $\Delta_x$ are positive, this means that \textit{positive triangularity and Shafranov shift contribute positively to} $B_{20}$, and thus we would expect MHD stability. There is a simple geometric explanation for this behaviour. Picture an increase of $\Delta_x$ as a bunching of cross sections on the outboard side of the configuration. As one goes from the magnetic axis outwards, each cross-section acquires more area on the inboard side compared to the outboard side. There $|\mathbf{B}|$ is larger, therefore, $B_{20}$ grows, and so does the magnetic well. Similarly, a positive triangularity brings the vertical turning points of the cross-section towards the inboard side, gathering a larger area on the high field side. Following this logic, any shaping that does not break left-right symmetry should not affect stability. 
\par
Although this geometric picture is simple, its link to stability is not as clear-cut as it may seem. When we deform the cross-sections by changing $\delta$ and $\Delta_x$ directly, the resulting equilibrium generally supports a different pressure gradient; formally, $p_2$ adjust self-consistently in the background to satisfy $p_2/2I_2^2=-(3+\eta^4)\Delta_x/(1+\eta^4)^2-\eta^5\delta/(1+\eta^4)^2+\dots$. To discuss stability most straightforwardly we shall keep $p_2$ constant, and thus make it explicit as in Eqs.~(\ref{eqn:B20pd})-(\ref{eqn:B20pD}). In such a scenario the expressions for $B_{20}$, Eqs.~(\ref{eqn:B20pd})-(\ref{eqn:B20pD}), and those for the Mercier criterion, Eqs.~(\ref{eqn:axisDMercpd})-(\ref{eqn:axisDMercpD}), are (up to a factor of $1/2$) the same as far as the effects of second-order shaping are concerned. The clear geometric picture we had for Eq.~(\ref{eqn:B20Dd}) is however lost. Positive triangularity and Shafranov shift no longer lead to an unequivocal increase in the magnetic well. Only vertically elongated cross-sections preserve the benefit of positive triangularity ($\eta<1$), while horizontally elongated ones do so for the Shafranov shift. The reason for this difference is the hidden response of the shaping in each one of these cases. To keep the pressure constant (see before), if one increases triangularity in a configuration, then the Shafranov shift should go in the negative direction to hold the same pressure gradient (in a sense counteracting the triangular shaping). This opposite contribution to $B_{20}$ makes (de)stabilising triangularity or Shafranov shift dominate in different situations. The stabilising effect of triangularity, for $\eta<1$, beats any destabilising influence of the Shafranov shift, which dominates when $\eta>1$. The exact balance results in Eqs.~(\ref{eqn:B20pd})-(\ref{eqn:B20pD}). 
\par
The form of Eq.~(\ref{eqn:axisDMercpd}) is consistent with \cite[Eq.~(12.89)]{freidberg2014}. We have learnt, though, that one must be careful at offering a simple picture to explain the behaviour of stability. The simple geometric view offered by Freidberg for Eq.~(\ref{eqn:B20Dd}) breaks down.\footnote{In \cite{freidberg2014}, the author uses a different notation to that presented here. For reference, $q_0=1/\iota_0$, $\kappa=1/\eta^2$, $\epsilon_\mathrm{Freid}=\eta\epsilon$ and $\beta_p=\kappa|p_2|/\iota_0^2(1+\kappa^2)$. The qualitative argument for the stability behaviour is related to the field lines spending longer on the good curvature region. That may be related to the growth in $|\mathbf{B}|$, as the good curvature region is the high field side. However, as we have seen, stability results from balancing the opposing behaviour of triangularity and Shafranov shift.} As most relevant cross-sections are elongated vertically in practice, \textit{positive triangularity (bean-shaping) contributes favourably to stability}. This aligns with the common wisdom of how shaping affects stability.
\par
As is well known, increasing the pressure in a configuration leads to a deepening of the magnetic well. Here, formally, this is described by the unavoidable increase of $B_{20}$ with $|p_2|$, Eqs.~(\ref{eqn:B20pd})-(\ref{eqn:B20pD}). The effect of pressure on stability is, even if $B_{20}$ increases, destabilising in the usual scenario in which triangularity is kept constant, Eq.~(\ref{eqn:axisDMercpd}). This leads to the well-known \cite{lortz_nuhrenberg_1978ballooning,freidberg2014} \textit{equilibrium $\beta$ stability limit}.\footnote{The $\beta$-limit occurs when increasing the pressure gradient and keeping triangularity fixed. This is a limit that is independent of the aspect ratio of the configuration.} The Shafranov shift plays a central role in setting this limit, as can be seen by the avoidance of the $\beta$-limit for $\eta<1$ when fixing $\Delta_x$, Eq.~(\ref{eqn:axisDMercpD}).
\par
The significance of any of the effects described is only relative to the effect of other terms in the Mercier criterion. This comparison depends on lower order parameter choices. This is especially true for what we call the `intrinsic contribution' to stability, a term that does not involve any second-order parameters directly. Its stabilising contribution grows with elongation in the horizontal direction, but it is deteriorated by current (opposite to the contribution by the pressure gradient). The behaviour with current can be understood considering the limit of very large $I_2$. In this limit, the tokamak effectively becomes a $Z$-pinch, whose instability grows like $I_2^2$ (see Ch.~11 in \cite{freidberg2014}), a classic result that in the circular-cross-section-limit becomes $D_\mathrm{Merc}\propto1-\iota_0^2$ \cite[Ch.~12]{freidberg2014}. 
\par
\begin{figure}
    \centering
    \includegraphics[width=0.8\textwidth]{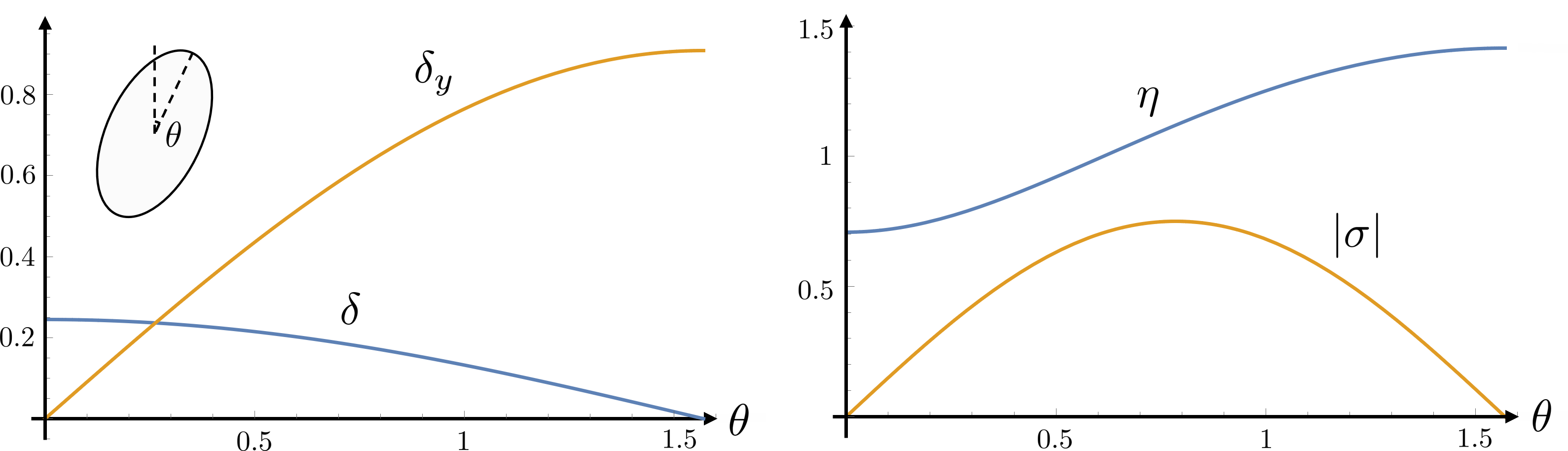}
    \caption{\textbf{Change in the stabilising triangularity effect with up-down symmetry breaking.} The plots show (left) the influence of regular and up-down triangularity on $B_{20}$, and (right) the $\eta$ and $\sigma$ parameters, as a function of the rotation of the ellipse, $\theta$. The elongation of the ellipse is kept constant, $\mathcal{E}=2$, and at $\theta=0$ is aligned with the vertical. }
    \label{fig:upDownBreaking}
\end{figure}

We pointed that any effect that did not break left-right symmetry at second order, namely $\delta_y$, could not affect $B_{20}$. But we had not touched upon the effect of up-down symmetry-breaking through $\sigma$. We shall spare the reader from the expressions one obtains in that case, which are not particularly illuminating. The procedure is, however, no different from the one we have adopted, as long as $\sigma$ is kept explicitly in the expressions. As pointed out in Appendix~\ref{sec:appTriang}, for a straightforward definition of shaping in that scenario, we redefine $\delta$, $\delta_y$ and $\Delta_x$ in the frame of the rotated ellipse. An example of how the effect of triangularity on stability changes with the up-down symmetry breaking is shown in Figure~\ref{fig:upDownBreaking}. 
As the ellipse rotates, the effect of triangularity $\delta$ shrinks, to the point of vanishing for $\vartheta=\pi/2$. At that point, $\delta$ represents vertical triangularity; and as we have seen, this has no effect on stability. The behaviour of $\delta_y$ in Fig.~\ref{fig:upDownBreaking} is the reverse to that of $\delta$. It goes from having no effect to having an effect which in this case surpasses the effect of $\delta$ at $\vartheta=0$. This difference in magnitude comes because the $\vartheta=0,~\pi/2$ cases are geometrically different. The major or minor axes are aligned with $X$ in each case, respectively.  

\section{Quasisymmetric stellarators}
The discussion of MHD stability and shaping above is a renewed outlook at a problem that has long been studied \cite{mercier1962,Solovev1970,lortz1976}. We confirmed that the behaviour in a tokamak aligns with the conventional wisdom of positive triangularity (in common elongated shapes) favouring stability, but the results are otherwise not new. The discussion above is, however, a valuable stepping stone towards dealing with the quasisymmetric problem. In quasisymmetry, like in axisymmetry, the Fourier coefficients of $|\mathbf{B}|$ in the near-axis expansion are constant, making most of the analysis the same. 

\subsection{Constancy of parameters}
Let us start by carefully considering the description of quasisymmetric configurations in the near-axis framework. By definition, the magnitude of the magnetic field in an equilibrium, quasisymmetric configuration (expressed in Boozer coordinates) is $|\mathbf{B}|=B(\psi,\chi=\theta-N\phi)$ \citep{boozer1983,helander2014,rodriguez2021gbc}. That is, the near-axis expansion of $|\mathbf{B}|$ is precisely analogous to that of axisymmetry. Thus, we expect to find a natural parametrisation of quasisymmetric configurations analogous to that of tokamaks. 
\par
To leading order, the shape of a magnetic axis should be chosen. For a quasisymmetric stellarator, any regular (i.e., with no vanishing curvature), closed space curve is valid, a priori, beyond a circle. The choice of axis fixes $N$, the direction of the symmetry, which corresponds to the self-linking number of the axis \citep{rodriguez2022phase}. The poloidal current is then $G_0=L/2\pi$ where $L$ is the total length of the axis. 
\par
At first order, elliptical shapes are described through $\eta$ and $\sigma$, like for the tokamak. However, in general $\sigma=\sigma(\phi)$ is a function of the toroidal angle; it is the solution to a periodic first-order Riccati equation (see Eq.~(A6) in \cite{garrenboozer1991b}, or Eq.~(2.14) in \cite{landreman2019}) with a choice of initial condition $\sigma(0)$ (vanishing in stellarator symmetry). Although this makes the shape of cross-sections in a quasisymmetric stellarator depend on $\phi$, the freedom at first order truly resides on two parameters, $\eta$ and $\sigma(0)$. This underlines the special nature of quasisymmetric stellarators.
\par
At second order in the distance from the magnetic axis we have the four parameters $p_2$, $B_{20}$, $B_{22}^C$ and $B_{22}^S$, the same number of constant parameters as in a tokamak. This is true in the ideal quasisymmetryic limit, which we will assume for the analysis in this section. However, in practice, one should not forget the limitations that arise through what has come to be known as the \textit{Garren-Boozer overdetermination problem} \citep{garrenboozer1991b}. Not every axis shape and parameter choice can consistently support quasisymmetry and equilibrium simultaneously through second order. Generally one is forced to relax the strict quasisymmetric requirement on, following \cite{landreman2019}, $B_{20}$, which becomes for consistency a function of the toroidal angle. Only a subset of near-axis choices \citep{rodriguez2022,landreman2022} have approximately constant $B_{20}$.\footnote{The lack of a unifying theory on the set of choices that conforms to quasisymmetry makes us refer to the axis shapes and their properties in most generality here. Only through some illustrating examples will the behaviour for the quasisymmetric subset be explored.} Thus, in practice, we can expect to find deviations between our idealised analysis and a more consistent one, driven by `errors' in quasisymmetry. For deviations in the range of $\Delta B_{20}\sim0.01$, one may estimate deviations in $\Delta V''\sim1$. When illustrating the findings in this section with practical examples, we will have to check that the idealised theory reproduces the actual near-axis configuration (see Appendix~\ref{sec:appQScheck}).

\subsection{Choosing a characteristic cross-section}
Although quasisymmetric configurations may be parametrised (for a given axis shape) by the same amount of parameters as a tokamak, flux-surfaces are naturally asymmetric. That is, the cross-section at each angle $\phi$, described analogously to the axisymmetric case\footnote{To draw the analogy, some elements need to be amended, such as $\eta\rightarrow\eta/\kappa$. It is important to realise that there is, however, an additional geometric effect that deforms cross-sections further in this more general case compared to the axisymmetric one. That deformation comes from the plane normal to a general magnetic axis, where cross-sections are particularly simple, not matching constant cylindrical angle planes (what may be referred to as the `lab-frame') where cross-sections are generally defined. The effect is a deformation of shapes, as described partially in Appendices~\ref{sec:appEllip}, \ref{sec:appTriang} and \ref{sec:appShafShift}. The details will appear in a future publication, but they do not affect in any significant way the discussion to follow.}, will be generally different. This makes the notions of ellipticity, triangularity and Shafranov shift functions of $\phi$. However, following the parametrisation of the axisymmetric scenario, we must be able to parametrise the whole configuration through the description of a single cross-section (and the axis shape). Given a cross-section and the axis shape, the remainder of the configuration then follows from the fulfilment of quasiymmetry and equilibrium. This simplicity is particular to quasisymmetric stellarators, but other optimised stellarators will also impose constraints on the shaping of their surfaces.
\par
{ In principle, one could consider the geometric features of any of the cross-sections as parameters.} In stellarator-symmetric stellarators, though, one cross-section is particularly simple: the up-down symmetric one. There are two such distinct cross-sections per field period, by stellarator symmetry occurring at $\phi=0,~\pi/N$, where $N$ is the number of field periods. We shall for simplicity focus on the cross-section at $\phi=0$. In common quasisymmetric configurations (see later section), this often corresponds to a characteristic bean-shaped cross-section, { and thus it is reasonable to consider it as representative in our discussion}. For such a cross-section, the normal vector of the Frenet-Serret frame points inwards along the major radius by construction\footnote{Under stellarator symmetry, $\phi\rightarrow-\phi$ and $z\rightarrow-z$, which requires the axis at the origin $\phi=0$ to be normal to $\hat{R}$.}, and the configuration presents a high degree of symmetry (curvature and torsion are even functions of $\phi$, and $\sigma$ is odd). The task is then to connect the features of this cross-section to stability.

\subsection{Shaping and MHD stability in a QS stellarator}
We relate the shaping of the up-down symmetric cross-section to stability in a form similar (conceptually) to how we approached the problem in the axisymmetric case. That is, we must first find a relation between $X_2$ and $Y_2$ (in this case at $\phi=0$) and typical shaping concepts, relate them to natural parameters of the near-axis framework, and finally draw the connection to the Mercier criterion. We will, for now, ignore the changes in shape of the cross-section that occur from the projection from the plane normal to the axis to the cylindrical coordinate system. We spare the reader from the algebra, and write the Mercier criterion in the following form,
 \begin{equation}
    \frac{\epsilon^2\pi^2 D_\mathrm{Merc}}{|p_2|G_0^2}=\mathcal{T}_{|p|}|p_2|+\mathcal{T}_\delta \delta+\Lambda, \label{eqn:DMercQSSplit}
\end{equation}
where all the interesting information lies in the forms of $\mathcal{T}$ and $\Lambda$. $\Lambda$, which includes the effects of the axis, $\eta$ and $\sigma$, and is important in determining the total stability of the configuration, is however not very illuminating (see Appendix~\ref{sec:appLambda}). Instead, we focus on the effect of triangularity, and see how the tokamak intuition and common wisdom claim holds.
\par
Using the appropriate relations, we obtain after significant algebra, 
\begin{equation}
    \mathcal{T}_\delta=3\eta\frac{(1-\alpha)+\bar{F}(1+\alpha)}{(3+\alpha)-\bar{F}(1+\alpha)}, \label{eqn:Td}
\end{equation}
where $\alpha=\eta^4/\kappa(0)^4$ and,
\begin{equation}
    \bar{F}=2\left[\frac{(I_2-\tau(0))/\kappa(0)^2}{\int_0^{2\pi}\mathrm{d}\varphi(I_2-\tau)/\kappa^2}\frac{\int_0^{2\pi}\mathrm{d}\varphi(1+\sigma^2+\eta^4/\kappa^4)}{1+\eta^4/\kappa(0)^4}-1\right],
    \label{eqn:Fbar}
\end{equation}
with $\tau$ the torsion of the axis. The expression in Eq.~(\ref{eqn:Fbar}) compares local quantities with their global average, acting as a measure of asymmetry in the stellarator. Readers familiar with the near-axis framework will recognise the averages as part of the expression for the rotational transform on axis, $\bar{\iota}_0=2G_0\eta^2\int[(I_2-\tau)/\kappa^2]/\int(1+\sigma^2+\eta^4/\kappa^4)$. In the axisymmetric limit, where the average and local quantities are the same, $\bar{F}\rightarrow0$, and $\mathcal{T}_\delta$ in Eq.~(\ref{eqn:Td}) reduces to Eq.~(\ref{eqn:axisDMercpd}), with $\alpha$ generalising $\eta\rightarrow\eta/\kappa(0)$. Thus, in this limit, the tokamak intuition holds: for a cross-section that is elongated in the vertical direction ($\alpha<1$), positive triangularity favours MHD stability. In contrast, negative triangularity contributes positively for $\alpha>1$.
\par
\begin{figure}
    \centering
    \includegraphics[width=\textwidth]{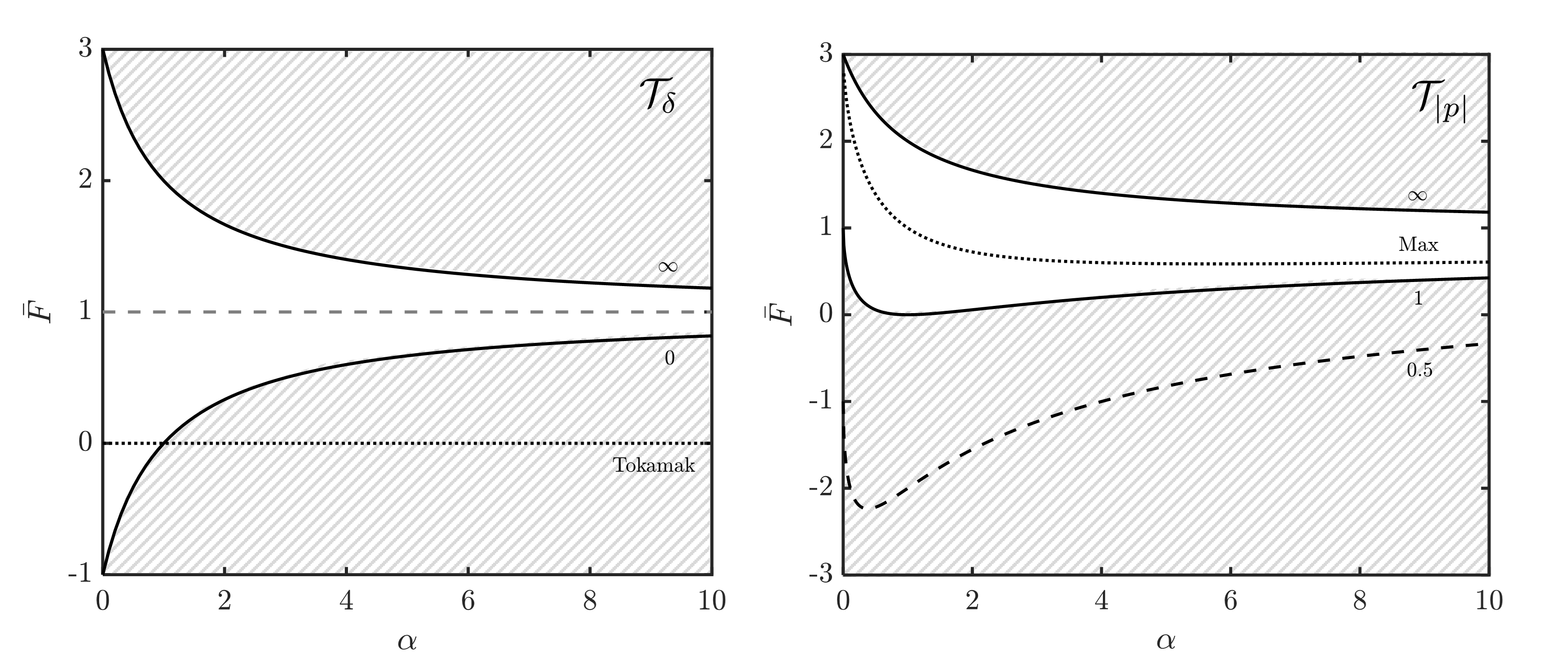}
    \caption{\textbf{Sign of $\mathcal{T}_\delta/3\eta$ and $\mathcal{T}_{|p|}\bar{\iota}_0^2/4\kappa^2$ in $(\alpha,~\bar{F})$ space.} The plots show $\mathcal{T}_\delta/3\eta$ (left) and $\mathcal{T}_{|p|}\bar{\iota}_0^2/4\kappa^2$ (right) in the $(\alpha,~\bar{F})$ space. The shaded region in the left plot represents the space for which $\mathcal{T}_\delta<0$, and thus positive triangularity is detrimental to the stability of the configuration. The shadow region for $\mathcal{T}_{|p|}$ also represents a negative sign corresponding to the destabilising effect of a pressure gradient. The dotted line on the left plot represents the case of the tokamak explored in the previous section, which shows the possibility of both triangularity signs being stabilising. The broken and dotted lines in the right plot correspond to the changing lower limit of the positively signed region as the magnitude of the geodesic contribution (the term with the integral $I$) is changed from maximal (dotted line) to half its magnitude from the axisymmetric limit (labelled 0.5). }
    \label{fig:FbaralpTdTp}
\end{figure}

The presence of $\bar{F}$ does not guarantee this behaviour in a general quasisymmetric stellarator. To understand the implications of this measure of asymmetry, we consider the representation of $\mathcal{T}_\delta/3\eta$ in $(\alpha,~\bar{F})$ space (see Figure~\ref{fig:FbaralpTdTp}). 
$\bar{F}$ changes the stabilising implications of triangularity drastically, especially in the $\bar{F}<0$ region. Any value $\bar{F}<(\alpha-1)/(1+\alpha)$ (of course, $\bar{F}<-1$) gives $\mathcal{T}_\delta<0$, and makes negative triangularity have a stabilising effect, contrary to common wisdom. In this region, though, $0>\mathcal{T}_d>-1$, and thus the effects of triangularity are moderate. To picture the meaning of negative $\bar{F}$, consider the limit of $\eta\sim0$. In that case, the first fraction in the square brackets of Eq.~(\ref{eqn:Fbar}) dominates $\bar{F}$, which for no toroidal current requires $\tau/\kappa^2$ to have a local minimum. 
\par
If this had a local maximum, then $\bar{F}>0$ and $\mathcal{T}_d$ would live in the upper portion of Fig.~\ref{fig:FbaralpTdTp}. In that case, we see that positive triangularity will benefit stability for moderate values of $\bar{F}$. Beyond $\bar{F}>(3+\alpha)/(\alpha+1)$ negative triangularity becomes once again stabilising, and in this case, strongly so. $\mathcal{T}_d$ shows a divergence at the boundary, corresponding to an unphysical Shafranov shift, indicating the break-down of the parametrisation chosen for the configuration. In the limit $|\bar{F}|\rightarrow\infty$, $\mathcal{T}_\delta\rightarrow-3\eta$.
\par
In summary, the common perspective on the contribution of triangularity (or bean shaping) to stability does not generally apply to quasisymmetric stellarators. For a significant asymmetry $|\bar{F}|$ the opposite is actually true. This is not a claim on the full MHD stability of a configuration, which we cannot make simply based on triangularity. Instead, it is a statement on the partial contribution of triangularity to the total MHD stability; that is, how a change in triangularity keeping the pressure, elliptic shaping, up-down symmetry and axis shape unchanged helps or worsens the stability of a configuration. Within the near-axis description this thought experiment has a precise formulation, and suggests that postive-triangularity, bean-shaped cross-sections do not necessarily improve stability.\footnote{Note that in practice, one must also keep QS, and thus the thought experiment can only be performed approximately. One may still talk about the contribution of triangularity to stability formally, as one may think of the limiting effect of a small change in $\delta$ (i.e., the derivative).} 
\par

\begin{figure}
    \centering
    \includegraphics[width=0.7\textwidth]{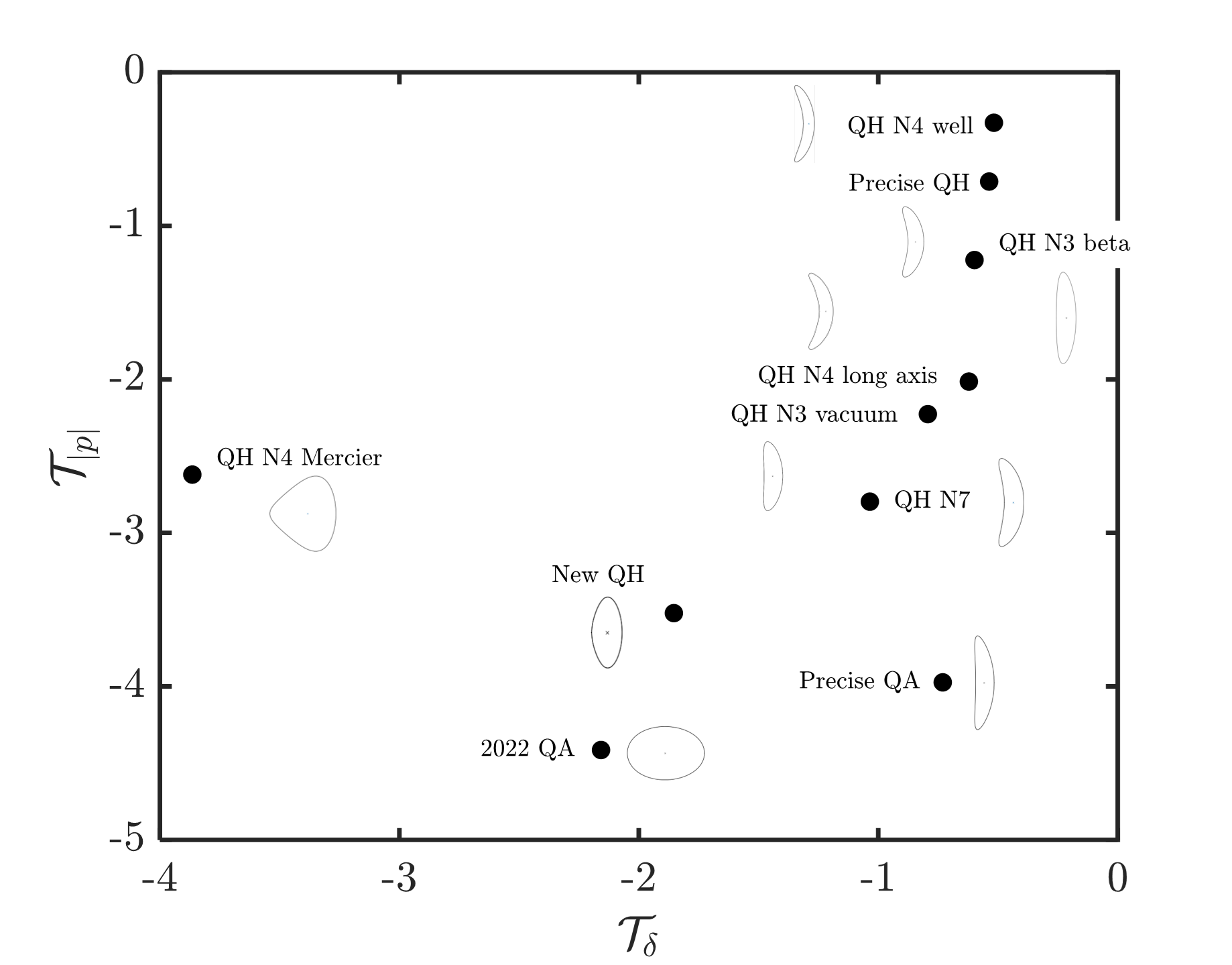}
    \caption{\textbf{Effect of triangularity and pressure on MHD stability for some quasisymmetric stellarators.} The plot shows as scatter points the factors regulating the effect of the triangularity ($\mathcal{T}_\delta$) and pressure gradient ($\mathcal{T}_{|p|}$) for several optimised quasisymmetric near-axis stellarators. The `precise' QA and QH are from \cite{landreman2021}, the new QH corresponds to the new optimised stellarator example from \cite{rodriguez2022}, while all others are from a recent publication \cite{landreman2022}. We chose those configurations with reduced $B_{20}$ variation so that the magnetic well computation, using a constant $B_{20}$, showed good agreement with the full $V''$. The cross-sections shown correspond to the $\phi=0$ cross-sections in each configuration.}
    \label{fig:shapeTriangPressMerc}
\end{figure}

\begin{table}
    \centering
    \begin{tabular}{c|c|c|c|c|c|c|c|c|c|c|c|}
         & PQA & PQH & NQH & 22QA & N3V & N4LA & N4W & N4M & N7 & N3B \\ \hline
        $\bar{F}$ & -3.1 & -1.4 & -1.6 & 5.6 & -1.5 & -1.3 & -1.7 & -1.6 & -0.8 & -1.6 \\
        $\delta$ & 4.9 & 4.6 & -1.7 & -0.3 & 0.9 & 1.2 & 11.8 & -9.1 & 1.3 & 0.9 \\
        $\mathcal{T}_\delta$ & -0.7 & -0.5 & -1.9 & -2.2 & -0.8 & -0.6 & -0.5 & -3.9 & -1.0 & -0.6 \\
        $V''/8\pi^2G_0$ & 1.1 & 1.1 & -1.4 & -0.2 & 1.9 & 2.0 & -0.5 & -11.9 & 7.3 & -1.7* \\
    \end{tabular}
\caption{\textbf{Details of the configurations in Figure \ref{fig:shapeTriangPressMerc}.} The table includes the values of $\bar{F}$, the triangularity $\delta$, the effect of triangularity $\mathcal{T}_\delta$, and the magnetic well $V''$ for the configurations represented in Figure~\ref{fig:shapeTriangPressMerc}. The short labels on top refer to PQA - precise QA, PQH - precise QH (from \cite{landreman2021}, NQH - new QH (from \cite{rodriguez2022}), 22QA - 2022 Qa, N3V - N3 vacuum, N4LA - N4 long axis, N4W - N4 well, N4M - N4 Mercier, N7 and N3B - N3 beta (all these from \cite{landreman2022}. For the latter instead of the magnetic well we show the $\epsilon^2 D_\mathrm{Merc}$, which shows that this finite $\beta$ configuration is unstable.}
    \label{tab:my_label}
\end{table}

What does this imply in practice? Are most of the existing cases aligning or misaligning with the common wisdom? From the analytic perspective, a definitive answer should explore the value of $\bar{F}$ in quasisymmetric configurations, but we shall content ourselves by presenting some examples of optimised near-axis quasisymmetric configurations, see Figure \ref{fig:shapeTriangPressMerc} and Table~\ref{tab:my_label}.\footnote{The configurations are described in the respective work, and we are here considering the cross-sections at $\phi=0$, stellarator symmetry points as defined in those. The cross-sections are shown in Fig.~\ref{fig:shapeTriangPressMerc}.} For all the cases analysed, remarkably, $\mathcal{T}_\delta<0$, a result that merits further investigation. As a result of this behaviour, the triangularity of the bean cross-section (see cross-sections in Fig.~\ref{fig:shapeTriangPressMerc}) is detrimental to MHD stability in most scenarios. Shaping contributes favourably only as an exception, of which the `new QH' \citep{rodriguez2022} and `QH N4 Mercier' \citep{landreman2022} configurations are two examples. The author in \cite{landreman2022} obtained the latter by optimising for quasisymmetry and a favourable Mercier criterion. Thus, finding the favourable contribution of triangular shaping is not surprising. The work here sheds some light on what appeared as a rarity in that paper. The `QH N4 well' \citep{landreman2022} presents a different scenario. This configuration presents a magnetic well, yet, according to our theory, it has the `wrong' triangular shaping. This is of course not inconsistent, as the triangularity contribution can be detrimental, yet the configuration remain stable. What is more surprising is the fact that the triangularity of this configuration is stronger compared to a similar configurations that was not optimised for a magnetic well. Optimising for stability seems to drive, in this case, the wrong shaping from the perspective of our theory. The disagreement is resolved by realising that on top of triangularity the axis shape was also modified. The change in the latter is sufficient to overcome the detrimental contribution of triangularity in this case. This scenario is reminiscent of the behaviour observed in optimisation of equilibrium boundaries such as in \citep{nuhrenberg1986,nuhrenberg2010}, examples on which the intuition on stability and shaping was built. Once again, the variation of the surface `triangularity' induces changes on the axis shape as well as other features in the equilibrium, explaining the potential disagreement. In the latter case a comparison to our theory is further complicated because of the variation of MHD stability properties throughout the volume and the necessary search for a near-axis description that adequately captures the behaviour of the global equilibrium near the axis. In any form, from the perspective of the near-axis analysis, the bean shaping we observe in most of the configurations in Fig.~\ref{fig:shapeTriangPressMerc} is detrimental and thus cannot be directly driven by the stability requirement, as it opposes it. It is natural to believe that it is the requirement of quasisymmetry or some other property that pushes the configuration to develop such bean-like feature. 
\par
Besides triangularity, the effect of pressure on stability is also modified in quasisymmetric configurations from that in tokamaks. In the latter, we unavoidably encountered the equilibrium stability $\beta$-limit (see Eq.~(\ref{eqn:axisDMercpd})). For a quasisymmetric stellarator, {we may write the following two equivalent forms,}
\begin{subequations}
\begin{align}
    \frac{\bar{\iota}_0^2\mathcal{T}_{|p|}}{4\kappa^2} &= \frac{2\alpha}{(3+\alpha)-\bar{F}(1+\alpha)}-\sqrt{\alpha}\mathcal{I}\left[\sqrt{\alpha},\sigma\right] \\
    &=-\frac{\alpha\left[(1-\sqrt{\alpha})^2-(1+\alpha)\bar{F}\right]}{(1+\sqrt{\alpha})[(3+\alpha)-(1+\alpha)\bar{F}]}-\sqrt{\alpha}\left(\mathcal{I}[\sqrt{\alpha},\sigma]-\frac{\sqrt{\alpha}}{1+\sqrt{\alpha}}\right),
    \label{eqn:TpExpress}
\end{align}
\end{subequations}
where $\mathcal{I}$ is the function introduced in Eq.~(\ref{eqn:DmIint}), and we have written the latter in a form in which the axisymmetric limit, Eq.~(\ref{eqn:axisDMercpd}), is straightforward (namely $\bar{F}\rightarrow0$ and $\mathcal{I}=\sqrt{\alpha}/(1+\sqrt{\alpha})$). The contribution of $\mathcal{I}$ is important, as it is negative and thus contributes to destabilising the stellarator for a finite plasma $\beta$. Its magnitude is bounded 
between $0<[\sqrt{\alpha}/(1+\sqrt{\alpha})]_\mathrm{min}\leq \mathcal{I}<1$, and thus one may take as orientative the case in which only the first fraction in Eq.~(\ref{eqn:TpExpress}) contributes to $\mathcal{T}_{|p|}$.
\par
The effect of $\bar{F}$, the measure of asymmetry, is shown in the right plot of Fig.~\ref{fig:FbaralpTdTp}. The factor $\mathcal{T}_{|p|}$ is no longer necessarily negative. Thus, an increase in the pressure gradient that improves stability is possible; naively, there would be no $\beta$ limit. This lack of a stability limit at zero magnetic shear in a stellarator is not a new concept, see \cite{hudson2005}. It could for instance lead to a configuration that is unstable at small plasma $\beta$, but becomes stable above some critical value, introducing the concept of a second stability regime. This would make reaching the finite $\beta$ equilibrium in practice difficult, but it is an attractive concept nonetheless. To picture $\mathcal{T}_{|p|}$, the integral $\mathcal{I}$ was largely simplified. Changing its contribution leads to changes in the region that satisfies $\mathcal{T}_{|p|}>0$. That region narrows when $\mathcal{I}$ is larger (tending towards the Max curve in the plot, which assumes $\mathcal{I}=1$) and widens when it becomes weaker (in the limit $I\sim0$ the lower bound going towards $\bar{F}\rightarrow-\infty$). In practice, the behaviour is more complex, as $\mathcal{I}$ and $\bar{F}$ (and even $\alpha$) are not completely independent. Figure~\ref{fig:shapeTriangPressMerc} shows $\mathcal{T}_{|p|}$ for some quasisymmetric designs, showing that in all cases there appears to be a $\beta$ stability limit. 

\section{Discussion and conclusions}
In this paper, we have addressed how the shaping of poloidal cross-sections is related to the MHD stability of toroidal plasmas in order to assess the common conception of positive triangularity bean shapes being favourable. We investigated axisymmetric tokamaks and quasisymmetric stellarators through a near-axis framework. 
\par
The analysis required constructing and defining conventional shaping notions such as triangularity and Shafranov shift within the inverse{-coordinate} near-axis-expansion framework. The work builds on \cite{landreman2019} and differs from the so-called direct approach by \cite{mikhailovskii1979} (or more recent efforts like \cite{kim2021}). The direct involvement of the magnetic field magnitude in our framework is convenient for discussing MHD stability beyond axisymmetry. 
\par
In the tokamak limit, we reproduce and more systematically explain existing results \cite{freidberg2014} on how shaping (especially triangularity) affects MHD stability through the Mercier criterion. Only for vertically elongated cross-sections, is positive triangularity MHD-stabilising. This dependence on elongation is a consequence of the contribution from the Shafranov shift, which wins over the destabilising effect of negative triangularity when cross-sections are horizontally elongated. The worsening of stability with increased pressure gradients and, thus, the appearance of a $\beta$ limit was also proven.
\par
We then explored the effects of shaping and pressure in stellarator-symmetric, quasisymmetric stellarators. We expressed the stability criterion in terms of the shape of a representative up-down symmetric cross-section, which together with an axis shape is sufficient to parametrise the whole configuration. The change of stability as one changes the shape of such cross-section was then studied. In practice, when the asymmetry in the problem is taken into consideration, we show that in most cases (including all the particular examples considered) negative triangularity is stabilising, contrary to current belief. The positive triangularity bean shapes most commonly encountered in QS stellarators thus appear to oppose stability (see Fig.~\ref{fig:shapeTriangPressMerc}), even when the configurations may be overall stable. The presence of these characteristic shapes must then correspond to a different property. Note that although we show this to hold for many existing optimised near-axis configurations, the behaviour is not necessary, and is possible to change it by tweaking the asymmetry measure $\bar{F}$, see Eq.~(\ref{eqn:Fbar}). 
\par
This added flexibility also exists for the effect of the pressure gradient. Unlike in the axisymmetric case, finite $\beta$ corrections can lead to a more MHD-stable configuration. Such behaviour requires particular choices of magnetic axis shapes and parameters, but is not seen in the examples considered. A deeper understanding of their feasibility requires future work on $\bar{F}$ and its relation to quasisymmetry and axis shapes. 
\par
This work suggests that the relation between cross-sections and stability in a quasisymmetric stellarator is complicated and does not conform necessarily to the need of a bean-shaped cross-section for stability. Making a general statement regarding the benefit of bean-shaped cross-sections to MHD stability in a general stellarator appears to be misleading. This is worth further exploration beyond quasisymmetry.

\appendix

\section{Mercier criterion} \label{sec:appMercier}
The Mercier criterion scalar $D_\mathrm{Merc}$ used in this paper can be written as $D_\mathrm{Merc}=D_s+D_w+D_d$ (page 23 in \cite{bauer2012} or \cite{landreman2020}),\begin{subequations}
    \begin{gather}
        D_s=\frac{1}{16\pi^2}\left(\frac{\mathrm{d}\iota}{\mathrm{d}\psi}\right)^2-\frac{1}{(2\pi)^4}\frac{\mathrm{d}\iota}{\mathrm{d}\psi}\int\mathrm{d}S\frac{\mu_0\mathbf{j}-I'\mathbf{B}}{|\nabla\psi|^3}, \\
        D_w=\frac{\mu_0}{(2\pi)^6}\frac{\mathrm{d}p}{\mathrm{d}\psi}\left(\frac{\mathrm{d}^2 V}{\mathrm{d}\psi^2}-\mu_0\frac{\mathrm{d}p}{\mathrm{d}\psi}\int\frac{dS}{B^2|\nabla\psi|}\right)\int\mathrm{d}S\frac{B^2}{|\nabla\psi|^3}, \\
        D_d=\frac{1}{(2\pi)^6}\left(\int\mathrm{d}S\frac{\mu_0\mathbf{j}\cdot\mathbf{B}}{|\nabla\psi|^3}\right)^2-\frac{1}{(2\pi)^6}\left(\int\mathrm{d}S\frac{B^2}{|\nabla\psi|^3}\right)\int\mathrm{d}S\frac{(\mu_0\mathbf{j}\cdot\mathbf{B})^2}{B^2|\nabla\psi|^3}.
    \end{gather} \label{eqn:DMercier}
\end{subequations}
We reproduce this form for reference-sake, as different forms of the Mercier criterion exist in the literature. In this case the instability criterion is $D_\mathrm{Merc}<0$. 

\section{Elliptic shaping} \label{sec:appEllip}
In this appendix, we derive in some more detail the expressions in Section \ref{sec:ellip} which relate the parameters $\eta$ and $\sigma$ of the near-axis framework to the geometric properties of the elliptic cross-sections to leading order. Let us start by writing down the ellipse equation in the canonical form of a second-order polynomial. To do so we define for the leading cross-section $X=\eta\cos\chi$ and $Y=(\sin\chi+\sigma\cos\chi)/\eta$, following Eq.~(\ref{eqn:ellipEq}). To construct the ellipse equation we seek expressions for $\sin\chi$ and $\cos\chi$, so that using the fundamental trigonometric relation of $\cos^2\chi+\sin^2\chi=1$, 
\begin{equation}
    X^2(1+\sigma^2)-2\eta^2\sigma XY+\eta^4Y^2=\eta^2.
\end{equation}
From this form, one may then construct the rotation angle and ellipticity of the ellipse,
\begin{subequations}
    \begin{gather}
        \mathcal{E}=\frac{F}{2\eta^2}\left[1+\sqrt{1-\frac{4\eta^4}{F^2}}\right], \\
        \tan\vartheta=\frac{F-2\eta^4}{2\sigma\eta^2}\left[1+\sqrt{1+\frac{4\sigma^2\eta^4}{(F-2\eta^4)^2}}\right],
    \end{gather} \label{eqn:elliptRot}
\end{subequations}
and $F=1+\sigma^2+\eta^4$. Here \textit{elongation} is defined as the ratio of the major to the minor radius, and $\vartheta$ is the angle between the major radius and the positive $X$ direction.
\par
The form of these expressions is reminiscent of a solution to a quadratic equation. In fact, one may rearrange Eqs.~(\ref{eqn:elliptRot}) in the following form,
\begin{subequations}
    \begin{gather}
        \mathcal{E}^2-\frac{1+\sigma^2+\eta^4}{\eta^2}\mathcal{E}+1=0, \\
        \Theta^2+\frac{\eta^4-1-\sigma^2}{\sigma\eta^2}\Theta-1=0,
    \end{gather}
\end{subequations}
where $\Theta=\tan\vartheta$. Rearranging the latter and using the double-angle formula, we obtain
\begin{equation}
    \tan 2\vartheta=\frac{2\Theta}{1-\Theta^2}=\frac{2\sigma\eta^2}{\eta^4-1-\sigma^2},\tag{\ref{eqn:ellipAng2}}
\end{equation}
as used in the text. In an analogous way, and defining $\mathcal{E}=\tan e$,
\begin{equation}
    \sin 2e=\frac{2\mathcal{E}}{1+\mathcal{E}^2}=\frac{2\eta^2}{1+\sigma^2+\eta^4}. \tag{\ref{eqn:ellipAng1}}
\end{equation}
We give the geometric interpretations of the angles $\vartheta$ and $e$ in Figure~\ref{fig:ellipseAng}. 
\par
We may also invert these relations to obtain a form for $\eta$ and $\sigma$ in the geometric $\vartheta$ and $e$. Using Eqs.~(\ref{eqn:ellipAng1})-(\ref{eqn:ellipAng2}),
\begin{subequations}
    \begin{gather}
        \sigma^2=\left(\frac{\sin 2\vartheta}{\tan 2e}\right)^2, \\
        \eta^2=\frac{1+\cos2\vartheta\cos2e}{\sin2e}.
    \end{gather}
\end{subequations}
Note that these expressions agree with what we know in the up-down symmetric limit, namely, that in the limit of the major radius being aligned with the curvature direction ($\vartheta\sim0$ and $\eta>1$), then $\eta^2=1/\mathcal{E}$, and when aligned along the binormal, $\eta^2=\mathcal{E}$.
\par
We remind the reader that the above is a description of the shape of the cross-section to leading order in the plane normal to the magnetic axis. In the tokamak case, this is what we shall call cross-sections in the `lab frame;, namely, poloidal cross-sections that result from cuts of the configuration at a constant cylindrical angle. In a more general stellarator, this shape in the normal plane is not the same as in the `lab frame'. This difference reduces to a projection factor that modifies the shape of the cross-section. These changes are important to consider in the context of, for instance, quasisymmetric stellarators. With this in mind, we must consider the reinterpretation of $\eta$ as $\eta\rightarrow\eta/\kappa$ and the projection factor.
\par
\begin{figure}
    \centering
    \includegraphics[width=0.3\textwidth]{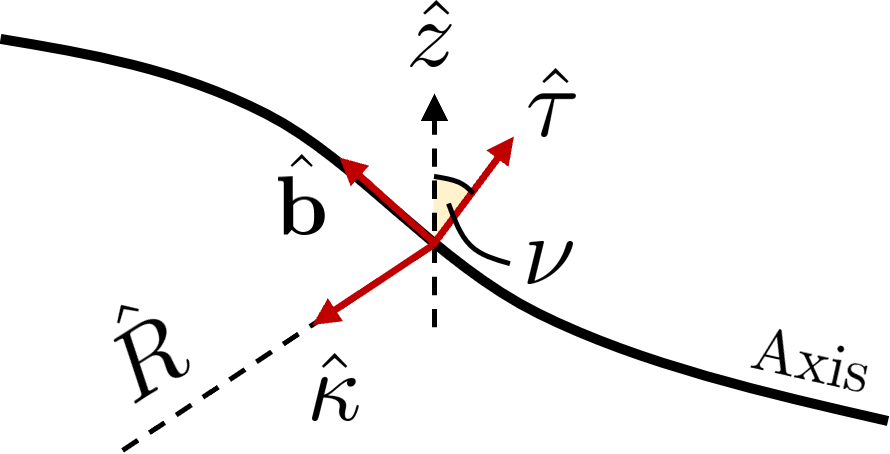}
    \caption{\textbf{Definition of the slant angle $\nu$.} Diagram showing the definition of the angle $\nu$ measuring the inclination of the magnetic axis at the origin ($\phi=0$) with the `lab' cylindrical coordinate system. The symbols have their usual meaning. }
    \label{fig:nuDef}
\end{figure}
Some attempts to describe the latter have been made by \cite{landreman2018a} and \cite{rodriguez2022}, but the geometric meaning is generally obscure and complicated. In a forthcoming paper, we will present a rigorous geometric form that describes these changes. For now, we content ourselves with understanding its primary consequences, focusing on the stellarator-symmetric, quasisymmetric stellarator, particularly the cross-section at $\phi=0$. The axis at this point has, by construction, its normal aligned with the major radius. However, generally, the magnetic axis is rotated about $\hat{R}$ by an angle $\nu$ (i.e., the angle between the binormal and the vertical $\hat{z}$). From this inclination, the cross-section in the plane normal to the axis is re-scaled $Y\rightarrow Y/\cos\nu$ in the `lab frame'. The ellipse thus shrinks in the vertical direction. For the most part, this is a simple adjustment that leaves $\eta$ and $\sigma$ with most of their geometric interpretation.

\section{Details on triangularity in the near-axis framework} \label{sec:appTriang}
In the main text, we assessed the effect of the shaping through second-harmonic modulation of the cross-sections. We did so in a geometrically intuitive form to motivate the relevant measure of the shaping strength and the final form of $\delta_\mathrm{tok}$. We did, however, not provide a derivation for the final expression for triangularity, Eq.~(\ref{eqn:triangAsymp}). Filling that gap is the purpose of this Appendix.
\par
\begin{figure}
    \centering
    \includegraphics[width=0.4\textwidth]{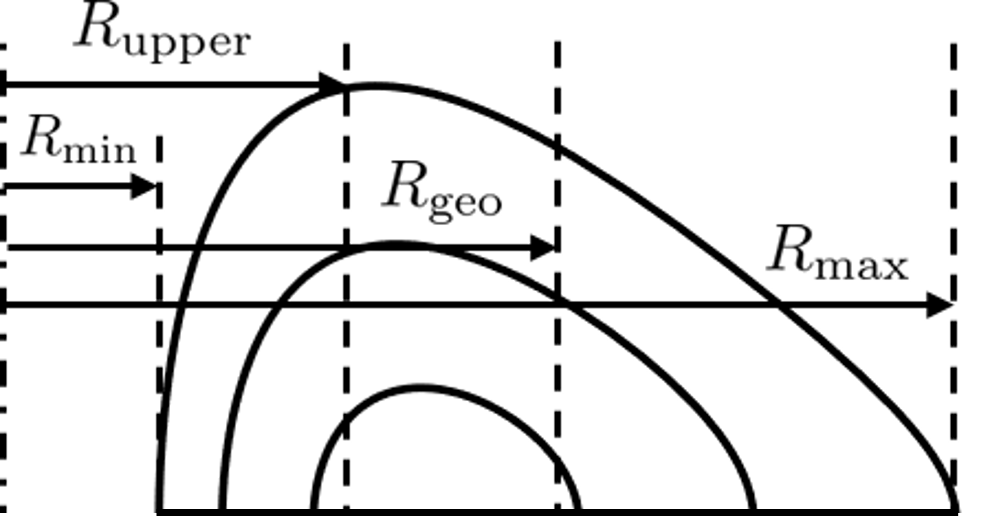}
    \caption{\textbf{Basic definitions for triangularity.} The diagram defines the length scales needed to compute the triangularity of an up-down symmetric cross-section indicated for the outermost surface). The evolution of the geometric centre $R_\mathrm{geo}$ from one flux surface to another defines the Shafranov shift, $\Delta_X$, with the negative sign indicating displacement outwards.}
    \label{fig:triang_diagram}
\end{figure}

Let us commence by defining triangularity in an up-down symmetric tokamak. We write this definition as $\delta_\mathrm{tok}=(R_\mathrm{geo}-R_{\mathrm{upper}})/a$, where $R_\mathrm{geo}=(R_\mathrm{min}+R_\mathrm{max})/2$, $a=(R_\mathrm{max}-R_\mathrm{min})/2$, $R_\mathrm{\mathrm{upper}}$ is the position of the turning point in the vertical direction, and $R_{\mathrm{min}/\mathrm{max}}$ are the leftmost and outermost points of the cross-section along the symmetry line\footnote{Note that these are not the same as the minimum and maximum radial positions when the triangularity is large enough to form a bean shape. In that case, we consider the positions along the symmetry line. This choice is unnecessary in most of the tokamak literature, as it rarely considers bean shapes.}. See the diagram in Fig.~\ref{fig:triang_diagram} for a depiction of these measures. Triangularity is the relative displacement of the vertical tips of the cross-section from the mid-point along the symmetry line. 
\par
To make rigorous contact with the near-axis modulation, consider $X$ and $Y$ (the normal and binormal directions of the cross-section) to coincide with the `lab frame'. In the tokamak context, this is, in fact, correct (taking into consideration the minus sign $X\rightarrow-R$ following the direction in which the major radius points), a notion that needs some adjustment in the more general quasisymmetric case, which we shall briefly touch upon at the end of this Appendix. Assuming up-down symmetry, and first only keeping $X_{22}^C$ and $X_{20}$ shaping, we may find the various geometric quantities in $\delta_\mathrm{tok}$ straightforwardly. The points $R_\mathrm{min}$ and $R_\mathrm{max}$ are $R_\mathrm{min/max}=1\mp \epsilon X_{11}^C-\epsilon^2(X_{20}+X_{22}^C)$. The turning point $R_\mathrm{upper}$ can simply be found by requiring $\partial_\chi Y=0$, which occurs at $\chi=\pi/2$ asymptotically, and thus $R_\mathrm{upper}=1-\epsilon^2(X_{20}-X_{22}^C)$. With these results, it then follows that,
\begin{equation}\label{eqn:triang_x2c}
    \delta_\mathrm{tok}=-2\epsilon\frac{X_{22}^C}{X_{11}^C},
\end{equation}
for a positive $X_{11}^C$. A positive value of $X_{22}^C$ then corresponds to what is known as negative triangularity (end-points of the cross-section pointing in the direction of larger $R$). We see the direct relation between $X_{22}^C$ and triangularity through the measure of strength motivated in the main text. 
\par
The effect of $Y_{22}^S$ on tokamak triangularity is analogous to that of $X_{22}^C$. In that case, we may once again compute $R_\mathrm{upper}$ as the position for the turning point $\partial_\chi Y=0$. Doing so yields an expression for $\cos\chi$ for the turning point, which may be expanded asymptotically in $\epsilon$, as $\cos\chi\sim2\epsilon Y_{22}^S/Y_{11}^S$. In this limit, $R_\mathrm{upper}\approx1+2\epsilon^2X_{11}^CY_{22}^S/Y_{11}^S$. With this and the symmetry line being unaffected, $\delta_\mathrm{tok}\sim2\epsilon Y_{22}^S/Y_{11}^S$. Importantly, the term is analogous to $X_{22}^C$, but with the opposite sign. 
\par
Of course, in general, these two forms of `triangular' shaping coincide and thus will interact in some form to result in a net triangularity. Here the asymptotic nature of the approach becomes highly valuable. In the limit $\epsilon\rightarrow 0$, each of these contributes to the total triangularity independently,\footnote{The change to the vertical turning point is of order $\epsilon$, and thus $X_{22}^C$ effects would only be affected at order $\epsilon^3$ (one order to high). Similarly, in reverse.}
\begin{equation}
    \delta_\mathrm{tok}\approx 2\epsilon\left(\frac{Y_{22}^S}{Y_{11}^S}-\frac{X_{22}^C}{X_{11}^C}\right). \tag{\ref{eqn:triangAsymp}}
\end{equation}
Before concluding this Appendix, we consider the case of triangularity in quasisymmetric stellarators. The derivation above holds at every toroidal angle $\phi$ in which the cross-sections are up-down symmetric.
\subsection{Projection to `lab-frame'} 
We expect the deformation of cross-sections when going from the plane normal to the axis to the cylindrical `lab-frame' to affect triangularity. After careful consideration of geometry and asymptotics, we find that the triangularity in the `lab frame' of the up-down symmetric cross-section at the origin of a quasisymmetric stellarator is
\begin{equation}
    \delta_\mathrm{lab}=\delta_\mathrm{tok}+\frac{\epsilon}{R_0}\left[\frac{1}{2}\left(\frac{\kappa}{\eta}\right)^3\left(1+3\frac{\partial_\phi^2 R_0}{R_0}\right)-\frac{\eta}{\kappa}\right]\sin^2\nu,
\end{equation}
where $\phi$ represents the cylindrical coordinate, $R_0$ is the radial position of the magnetic axis, $\nu$ the angle defined previously denoting the deviation of the axis binormal from $\hat{z}$, and all quantities are evaluated at the origin $\phi=0$. Of course, in the limit of $\nu=0$, the triangularity is precisely that computed before, $\delta_\mathrm{tok}$. 
\par
The critical realisation is that the only difference is in an order $\sin^2\nu$ term that shifts the value of $\delta_\mathrm{tok}$. The difference depends solely on the shape of the axis and $\eta$, but no other higher-order quantity. A change of triangularity in the normal plane thus directly leads to an equivalent change in the `lab frame'. No rescaling occurs because triangularity is a quantity normalised to the underlying ellipse shape, which we also learnt to be deformed. The ratio remains unchanged. Thus, with this caveat, we may mostly ignore this difference when discussing the effects of increasing or decreasing triangularity, understood as a consequence of second-order choices. 
\subsection{Triangularity and up-down asymmetry} 
We constructed the notion of triangularity in the context of an up-down symmetric cross-section. Breaking this symmetry needs some adjustments in the analysis. The symmetry in the near-axis framework may be broken in two different ways. On the one side, asymmetry could arise purely at second order from the modulation of $X_{22}^S$ and $Y_{22}^C$. As the main text argues, their effect is analogous to triangularity but in the vertical direction. Thus one may define $\delta_y$ (the vertical triangularity) as a measure of asymmetry.
\par
The up-down symmetry may also be broken at first order whenever the elliptical cross-sections are not aligned with the Frenet-Serret basis. In that case, regular and up-down triangularity will no longer correspond to the expressions for $\delta_\mathrm{tok}$ and $\delta_y$. The rotation of the underlying ellipse mixes the shaping harmonics into a new linear combination. A reasonable way to define the geometry of such shapes is to define $\delta_\mathrm{tok}$ and $\delta_y$ not in the Frenet frame but rather in the frame of the ellipse. This change requires a mapping of the shape coefficients that takes the rotation of the ellipse $\vartheta$ into account. Defining $X'$ and $Y'$ as the rotated coordinates, and $\mathbb{C}=\cos\vartheta$ and $\mathbb{S}=\sin\vartheta$, where $\vartheta$ is the rotation angle of the ellipse as given by Eq.~(\ref{eqn:ellipAng2}), we rotate the original ellipse and define a new poloidal angle $\chi'$ so that,
\begin{equation}
    \begin{pmatrix}
        X'\\
        Y'
    \end{pmatrix}=\frac{1}{\eta}\begin{pmatrix}
        \sqrt{\mathbb{S}^2+(\eta^2\mathbb{C}+\sigma\mathbb{S})^2}\cos\chi' \\
        \sqrt{\mathbb{C}^2+(\eta^2\mathbb{S}+\sigma\mathbb{C})^2}\sin\chi'
    \end{pmatrix}.
\end{equation}
This expression describes a frame-aligned ellipse, a baseline we used to define triangularity in this Appendix. Here $\chi'=\chi-\Theta$, where $\tan\Theta=-\mathbb{S}/(\eta^2\mathbb{C}+\sigma\mathbb{S})$. We get a similar transformation for the higher order. That is, we rotate the $X_2$ and $Y_2$ components by $-\vartheta$ and re-express the harmonics in $\chi=\chi'-\Theta$ to obtain $X_2'$ and $Y_2'$. Then we define using Eqs.~(\ref{eqn:triangAsymp}) and (\ref{eqn:triangYAsymp}), $\delta'$ and $\delta_y'$, which will involve generally complicated linear combinations of second-order parameters. Doing so is algebraically untidy but may be accomplished using computational algebra. We shall not write down the expressions here, as they do not provide much insight other than showcasing the mixing effect of the ellipse rotation. We only use some numerical examples of it in the main text.

\section{Details on the generalised Shafranov shift}\label{sec:appShafShift}
We introduced in the main text a definition of a generalised Shafranov shift that describes the relative displacement of cross-section centres. This generalised form was originally presented in \cite{rodriguez2022}. The expression reduces to the axisymmetric limit as shown by \cite{landreman2020}. As emphasised, the arbitrariness to the centre of cross-sections extends to the definition of the Shafranov shift. This Appendix will motivate the form in Eq.~(\ref{eqn:shafShiftNAE}) taken as the definition of the Shafranov shift. 
\par
Consider a coordinate map that maps the cross-sections in the plane normal to the axis (however complicated) into circular shapes (a sort of \textit{normal form} of the cross-section). Once we have performed such a mapping, we end up with circles, which have a unique centre. The transformation will generally be complicated, but it must exist, as the cross-sections are, after all, an embedding of $S^1$ in the plane. 
\par
Let us state how to achieve this at first order. The main idea will be to cast the equations describing $X$ and $Y$ in a form that explicitly gives $\cos\chi$ and $\sin\chi$. In matrix form,
\begin{equation}
    \begin{pmatrix}
        X \\
        Y    
    \end{pmatrix}=\epsilon\begin{pmatrix}
        X_{11}^C\cos\chi \\
        Y_{11}^S(\sin\chi+\sigma\cos\chi)
    \end{pmatrix}=\epsilon\underbrace{\begin{pmatrix}
        X_{11}^C & 0 \\
        \sigma Y_{11}^S & Y_{11}^S
    \end{pmatrix}}_{\mathcal{M}_{(1)}}\begin{pmatrix}
        \cos\chi \\
        \sin\chi
    \end{pmatrix}.
\end{equation}
Inverting this,
\begin{equation*}
    \epsilon\begin{pmatrix}
        \cos\chi \\
        \sin\chi
    \end{pmatrix}=\begin{pmatrix}
        1/X_{11}^C & 0 \\
        -\sigma/X_{11}^C & 1/Y_{11}^S
    \end{pmatrix}\begin{pmatrix}
        X \\
        Y    
    \end{pmatrix},
\end{equation*}
and computing the norm provides the equation of a circle $(X')^2+(Y')^2=\epsilon^2$, where
\begin{gather*}
    X' = \frac{X}{X_{11}^C}=\bar{X}, \\
    Y' = \frac{Y}{Y_{11}^S}-\sigma\frac{X}{X_{11}^C}=\bar{Y}-\sigma\bar{X}.
\end{gather*}
Note that this is the same as before when defining the ellipticity and rotation angles in the previous Appendix. The ellipses become circles in the transformed space $(X',Y')$.
\par
With this leading-order procedure in mind, we may construct the `circles' for the second-order shaping. To do so, we shall use trigonometric relations of the form $\cos2\chi=1-2\sin^2\chi$, to $O(\epsilon^2)$. With that, and multiplying through $\mathcal{M}^{(1)}$,
\begin{multline*}
    \begin{pmatrix}
        X' \\
        Y'
    \end{pmatrix}-\epsilon^2\begin{pmatrix}
        \bar{X}_{20}+\bar{X}_{22}^C \\
        \bar{Y}_{20}(1-\sigma)+\bar{Y}_{22}^C-\sigma\bar{X}_{22}^C
    \end{pmatrix}=\\
    \epsilon\begin{pmatrix}
        1+2\epsilon\bar{X}_{22}^S\sin\chi & -2\epsilon\bar{X}_{22}^C\sin\chi \\
        2\epsilon(\bar{Y}_{22}^S-\sigma\bar{X}_{22}^S)\sin\chi & 1+2\epsilon(\sigma\bar{X}_{22}^C-\bar{Y}_{22}^C)\sin\chi
    \end{pmatrix}\begin{pmatrix}
        \cos\chi \\
        \sin\chi
    \end{pmatrix},
\end{multline*}
where the overline indicates normalisation with respect to $X_{11}^C$ or $Y_{11}^S$ respectively. 
\begin{figure}
    \centering
    \includegraphics[width=0.6\textwidth]{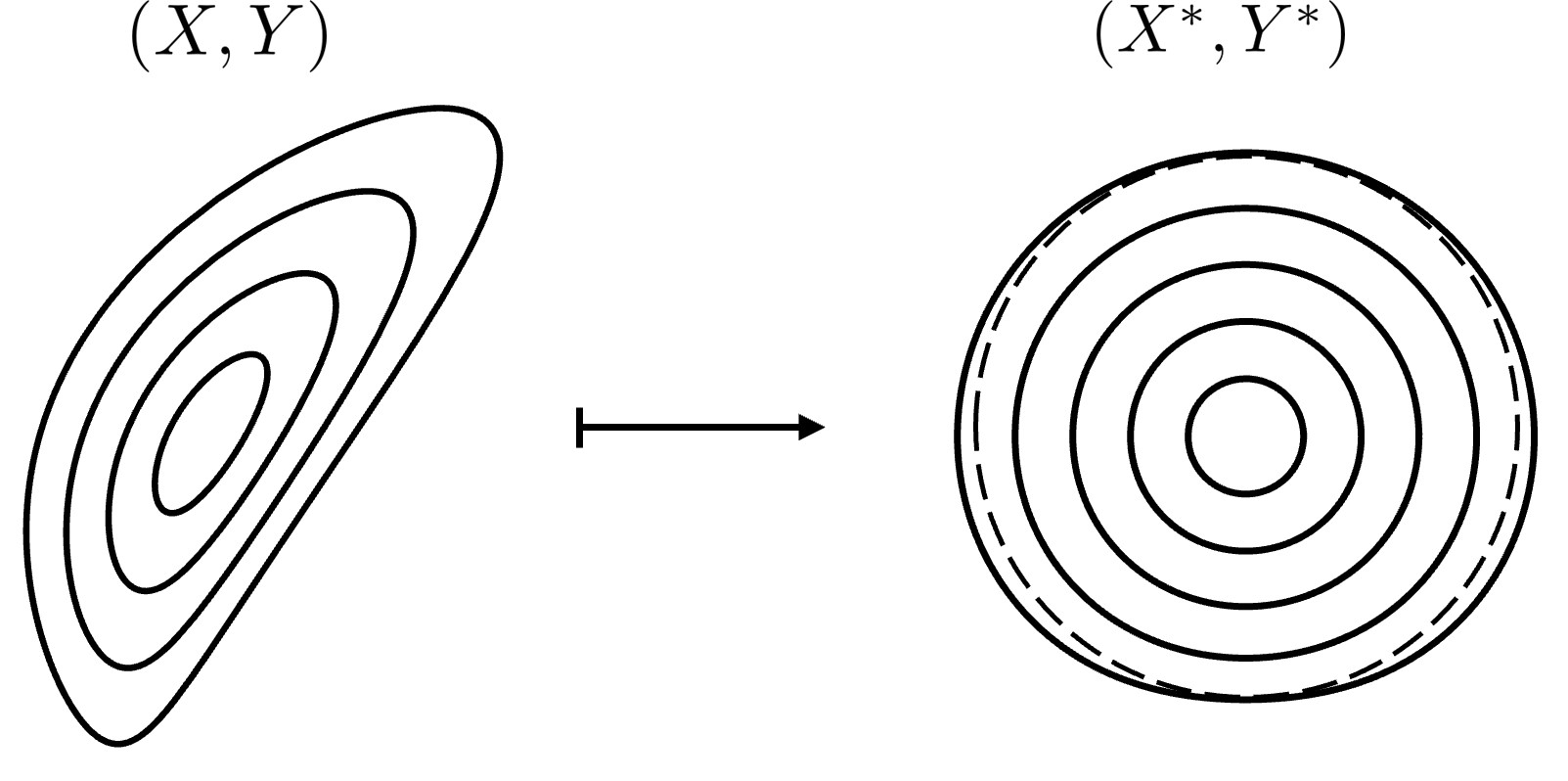}
    \caption{\textbf{Example of the coordinate map to define the Shafranov shift.} Example showing the coordinate map transformation of a second-order shape onto concentric circles. We can obtain and identify the cross-section shift from this mapping with the Shafranov shift. The deviation in $(X^*,Y^*)$ from a circle (broken line) results from higher-order effects.}
    \label{fig:mapCoordShaf}
\end{figure}
Inverting the matrix and keeping the relevant orders in $\epsilon$, we can define a circle $(X^*)^2+(Y^*)^2=\epsilon^2$,
\begin{multline}
    \begin{pmatrix}
        X^* \\
        Y^*
    \end{pmatrix}=\\
    =\frac{1}{1+2\epsilon(\bar{X}_{22}^S+\sigma\bar{X}_{22}^C-\bar{Y}_{22}^C)}\begin{pmatrix}
        1-2\epsilon\bar{Y}_{22}^C\sin\chi & 2\epsilon\bar{X}_{22}^C\sin\chi \\
        -(\sigma+2\epsilon\bar{Y}_{22}^S\sin\chi) & 1+2\epsilon\bar{X}_{22}^S\sin\chi
    \end{pmatrix}\begin{pmatrix}
        \bar{X}-\epsilon^2(\bar{X}_{20}+\bar{X}_{22}^C) \\
        \bar{Y}-\epsilon^2(\bar{Y}_{20}+Y_{22}^C)
    \end{pmatrix}.
\end{multline}
\par
Figure~\ref{fig:mapCoordShaf} shows an example of this transformation. The transformation matrix reduces to the ellipse map to leading order $\epsilon$. However, it is clear from this map that cross-sections have a relative shift. From that, we may read off the Shafranov shift,
\begin{subequations}
\begin{gather}
    \Delta_x=X_{20}+X_{22}^C, \tag{\ref{eqn:shafShiftNAEX}} \\
    \Delta_y=Y_{20}+Y_{22}^C. \tag{\ref{eqn:shafShiftNAEY}}
\end{gather}
\end{subequations}
We refer to these as \textit{generalised Shafranov shift}. It satisfies the necessary circular cross-section axisymmetric limit and holds for any $\phi$ and second-order shaping. The quantity $\Delta_x$ is directly related to the mid-point of the cross-section around the up-down symmetry line. The shift in the $Y$ direction has a similar meaning but in the perpendicular direction. We emphasise that the approach presented is not unique, and the transformation $(X,Y)\rightarrow (X^*,Y^*)$ could have been chosen in another way. However, those forms would not generally match the axisymmetric limit.\footnote{An example of this occurs if we express the matrix representing the linear transformation in terms of sines. To do so, we must use the corresponding form of the double-angle formulas, and we end up getting a shift $X_{20}-X_{22}^C$, which makes no geometric sense.} 
\par
To conclude this Appendix, we must briefly touch on the effect of the projection to the `lab frame' on the Shafranov shift. Focusing on the shift $\Delta_x$, relevant for the up-down symmetric cross-section, one may show that it remains invariant. Although effectively $X_{20}$ and $X_{22}^C$ each have a shift from the projection, these are opposed, and thus the Shafranov shift is invariant. No scaling is involved because $X$ is aligned with the major radius. Thus we expect to find a change on $\Delta_y$. These changes are particularly complex, as not only does the projection involve the $1/\cos\nu$ scaling, but the components of $Z_2$ are also involved. Thankfully we do not need to consider this.   

\section{Governing near-axis equations} \label{sec:equationsNAE}
In order to relate the near-axis shaping to $|\mathbf{B}|$-harmonics and other natural near-axis elements, it is necessary to know the expressions that relate them. These come from the asymptotic expansion in powers of the distance from the magnetic axis of the magnetic field and its governing equilibrium and magnetic equations. The original work by \cite{garrenboozer1991a,garrenboozer1991b} and \cite{landreman2019} are good places for reference of these equations, while \cite{rodriguez2020i} gives a more general form of the description beyond equilibria with isotropic pressure. We here use the notation in \cite{landreman2019}.
\par
\par
For completeness, we write down the expressions for the quasisymmetric case (which includes axisymmetry as a particular case). Following the notation of Eq.~(\ref{eqn:fExpansion}), the expansion components of position functions have the form,
\begin{equation}
    X = \epsilon X_{11}^C\cos\chi+\epsilon^2(X_{20}+ X_{22}^C\cos2\chi+X_{22}^S\sin2\chi),
\end{equation}
and similarly for $Y$ and $Z$.
\par
The $X_2$ components come from the requirements on $|\mathbf{B}|$ and the Jacobian (see Eqs.~(A34)-(A36) in \cite{landreman2019} or Appendix C in \cite{rodriguez2020i}),
    \begin{subequations}
    \begin{multline}
        X_{2,0}=\frac{1}{4 l' \kappa }\left[l' \kappa ^2 X_{1,1}^C{}^2+(l'\tau^2+\iota _0^2/l') \left(X_{1,1}^C{}^2+Y_{1,1}^C{}^2+Y_{1,1}^S{}{}^2\right)-2 \tau  Y_{1,1}^C X_{1,1}^C{}'+\right.\\
        \left.+2 \tau  X_{1,1}^C Y_{1,1}^C{}'+2 (\iota _0/l') \left(2 l' \tau  X_{1,1}^C Y_{1,1}^S{}+Y_{1,1}^S{} Y_{1,1}^C{}'-Y_{1,1}^C Y_{1,1}^S{}'\right)+\right.\\
        \left.+X_{1,1}^C{}'{}^2+Y_{1,1}^C{}'{}^2+Y_{1,1}^S{}'{}^2+4 Z_{20}'\right]\\
        +\frac{1}{ B_0^3 (l')^2 \kappa }\left[G_0^2 \left(B_{20}-\frac{3}{4}  B_0 \eta ^2\right)- B_0 G_0 (G_1+\iota_0I_1)\right], \label{eqn:X20}
    \end{multline}
    \begin{multline}
        X_{2,2}^C=\frac{1}{4\kappa}\left[4\frac{B_{22}^C}{B_0}-3\eta^2+\kappa^2(X_{11}^C)^2+(\tau^2-(\bar{\iota}_0/l')^2)\left((X_{11}^C)^2+(Y_{11}^C)^2-(Y_{11}^S)^2\right)+\right.\\
        \left.+\frac{1}{(l')^2}\left({(X_{11}^C)'}^2+{(Y_{11}^C)'}^2-{(Y_{11}^S)'}^2\right)-\frac{2\tau}{l'} Y_{11}^C{X_{11}^C}'+\frac{2\tau}{l'} X_{11}^C{Y_{11}^C}'+\right.\\
        \left.+\frac{2\bar{\iota}_0}{(l')^2}\left(Y_{11}^S{Y_{11}^C}'+Y_{11}^C{Y_{11}^S}'\right)+\frac{8\bar{\iota}_0}{l'}Z_{22}^S+\frac{4}{l'}{Z_{22}^C}'\right],
    \end{multline}
    \begin{multline}
        X_{2,2}^S=\frac{1}{2\kappa}\left[2\frac{B_{22}^S}{B_0}+(\tau^2-(\bar{\iota}_0/l')^2)Y_{11}^CY_{11}^S-\frac{\tau}{l'}\left(Y_{11}^S{X_{11}^C}'-X_{11}^C{Y_{11}^S}'\right)+\right.\\
        \left.+\frac{1}{(l')^2}{Y_{11}^S}'{Y_{11}^C}'-\frac{\bar{\iota}_0}{(l')^2}\left(X_{11}^C{X_{11}^C}'+Y_{11}^Y{Y_{11}^C}'-Y_{11}^S{Y_{11}^S}'\right)+\frac{4\bar{\iota}_0}{l'}Z_{22}^C+\frac{2}{l'}{Z_{22}^S}'\right],
    \end{multline}
    \end{subequations}
    where we define the pressure gradient to include the constant factor $\mu_0=4\pi\times 10^{-7}$ often shown explicitly, and we use the shorthand $l'=\mathrm{d}l/\mathrm{d}\phi$. Here (see Eqs.~(A27)-(A29) in \cite{landreman2019} or Eq.~(24) in \cite{rodriguez2020i}),
\begin{subequations}
        \begin{gather}
            {Z}_{20}=\frac{1}{4\eta^2\kappa^3l'}\left[\eta^4\kappa'-\kappa^4\left(\kappa'(1+\sigma^2)+\kappa\sigma\sigma'\right)\right], \\
            Z_{2,2}^S=\frac{1}{4\eta^2\kappa^2l'}\left[\eta^4\bar{\iota}_0+\kappa^3\left(\bar{\iota}_0\kappa(\sigma^2-1)-2\sigma\kappa'-\kappa\sigma'\right)\right], \\
            Z_{2,2}^C=\frac{1}{4\eta^2\kappa^3l'}\left[-2\bar{\iota}_0\kappa^5\sigma+\eta^4\kappa'-\kappa^4\left(\kappa'(\sigma^2-1)+\kappa\sigma\sigma'\right)\right],
        \end{gather}
    \end{subequations}
    which follow from the divergenceless and flux surface nature of the field. The shapes $\{X_2\}$ and the magnetic field harmonics $\{B_2\}$ are intimately related. 
    \par
    Then (see Eqs.~(A25)-(A26) in \cite{garrenboozer1991b} or Eqs.~(27)-(28) in \cite{rodriguez2020i}),
    \begin{subequations}
    \begin{gather}
        \Tilde{Y}_{2,2}^C=\frac{\kappa^2}{\eta^2}\left[\left(X_{2,2}^C-\Tilde{X}_{2,0}\right)\sigma+X_{2,2}^S\right], \\
        Y_{2,2}^S=-\frac{\kappa}{2}-\frac{\kappa}{\eta^2}\left[\left(X_{2,2}^C+\Tilde{X}_{2,0}\right)-X_{2,2}^S\sigma\right],
    \end{gather}
    \end{subequations}
    where $Y_{2,2}^C=Y_{2,0}+\Tilde{Y}_{2,2}^C$, and (see \cite{rodriguez2020i,rodriguez2020ii} in the isotropic limit),
    \begin{multline}
        Y_{20}=\frac{1}{2 \bar{\iota}_0 \kappa^2}\left[4l'\eta^2 \left(\tilde{Y}_{22}^C Z_{22}^S+Y_{22}^S \left({Z}_{20}-Z_{22}^C\right)\right)+\eta ^2 \tilde{Y}_{22}^C \left(2\frac{I_2}{B_0}-\tau\right) l'+\right.\\
        \left.+\eta ^2 {X_{22}^C}'
        -\kappa l'Z_{20} \left(\eta ^2-4 \kappa \left(X_{22}^C-\sigma X_{22}^S\right)\right)-4 l'\kappa^2 \sigma X_{22}^C Z_{22}^S-4 l'\kappa^2 X_{20} Z_{22}^C+\right.\\
        \left.+4l' \kappa^2 \sigma X_{20} Z_{22}^S+2 \eta ^2 \iota _0 X_{22}^S+4 l'\kappa^2 \sigma X_{22}^S Z_{22}^C+\eta ^2 \kappa l' Z_{22}^C-
        \eta ^2 X_{20}{}'+\right.\\
        \left.+\kappa^2 \left((\sigma X_{20} - \sigma X_{22}^C- X_{22}^S)\left(2\frac{I_2}{B_0}-\tau\right)l'+ \sigma \tilde{Y}_{22}^C{}'-2 \iota _0 (\tilde{Y}_{22}^C-\sigma Y_{22}^S)+ Y_{22}^S{}'\right)\right]
    \end{multline}
        \par
    Finally we have the self-consistent equilibrium condition which in an ideally quasisymmetric case reads (see \cite{thesis}), using the shorthand $\Tilde{\tau}=\tau-I_2/B_0$
        \begin{equation}
        4\mathcal{C}\left(-\frac{B_{20}}{B_0}+\frac{3\eta^2}{4}\right)+\mathcal{D}=0, \label{eqn:B20eqApp}
    \end{equation}
    where 
    \begin{subequations}
    \begin{gather}
        \mathcal{C}=-\frac{1}{G_0\eta^3\kappa^2}\left[\bar{\iota}_0\left(\kappa^4(1+\sigma^2)-3\eta^4\right)-4l'\eta^2\kappa^2\Tilde{\tau}\right],
    \end{gather}
    \begin{equation}
        \mathcal{D}= \frac{\mathcal{D}_{-1}}{\bar{\iota}_0}+\mathcal{D}_0+\bar{\iota}_0 \mathcal{D}_1+\bar{\iota}_0^2 \mathcal{D}_2+\bar{\iota}_0^3 \mathcal{D}_3, \label{eqn:DofB20}
    \end{equation}
\end{subequations}
where,
\begin{subequations}
    \begin{equation}
        \mathcal{D}_{-1}=-\frac{8\kappa^2}{G_0\eta}\frac{\mathrm{d}}{\mathrm{d}\phi}\left\{\frac{1}{\kappa^2}\frac{\mathrm{d}}{\mathrm{d}\phi}\left[\tilde{\tau}^2+\frac{1}{(l')^2}\left(\frac{\kappa'}{\kappa}\right)^2+\frac{\kappa^2}{4}\right]\right\},
    \end{equation}
    \begin{multline}
        \mathcal{D}_0=\frac{48B_{22}^S}{B_0^2}\left(\frac{\sigma\tilde{\tau}}{\eta}+\frac{\eta\kappa'}{l'\kappa^3}\right)+\frac{48\tilde{\tau}}{B_0\eta}\left(-\frac{B_{22}^C}{B_0}+\frac{3\eta^2}{4}\right)+\frac{16B_{\alpha1}\tilde{\tau}}{\eta B_{\alpha0}B_0}+\frac{8\eta}{\kappa^4B_0}\left[-4\kappa^4\tilde{\tau}+\right.\\
        \left.+3\kappa^2\tilde{\tau}^3+8\tilde{\tau}\left(\frac{\kappa'}{l'}\right)^2-\frac{\kappa}{(l')^2}(-9\kappa'\tilde{\tau}'+5\tilde{\tau}\kappa'')-2\kappa^2\frac{\tilde{\tau}''}{(l')^2}\right]+\\
        +\frac{4\sigma}{G_0\eta\kappa^4}\left[-3\kappa^5\kappa'+24\kappa\frac{(\kappa')^3}{(l')^2}-10\kappa^4\tilde{\tau}\tilde{\tau}'-30\kappa^2\kappa'\frac{\kappa''}{(l')^2}-2\kappa^3\left(\tilde{\tau}^2\kappa'-2\frac{\kappa^{(3)}}{(l')^2}\right)\right]+\\
        +\frac{2I_2^2}{B_0 \eta\kappa^2} \left(-\eta^2 \tilde{\tau} +\sigma \kappa \frac{\kappa'}{l'}\right)+\frac{4\eta I_2}{B_0 \kappa^4} \left(- \kappa^4 + 2  \kappa^2 \tilde{\tau}^2 + 2 \frac{(\kappa')^2}{(l')^2}\right),
    \end{multline}
    \begin{multline}
        G_0\mathcal{D}_1=12\left(-\frac{B_{22}^C}{B_0}+\frac{3\eta^2}{4}\right)\left(\frac{3\eta}{\kappa^2}+\frac{\kappa^2}{\eta^3}(1-\sigma^2)\right)+\frac{4B_{\alpha1}}{G_0}\left(\frac{3\eta}{\kappa^2}-\frac{\kappa^2}{\eta^3}(1+\sigma^2)\right)+\frac{24\kappa^2\sigma}{\eta^3}\frac{B_{22}^S}{B_0}-\\
        -\frac{1}{\eta\kappa^6}\left[4\kappa^4\left(2\kappa^4+5\eta^4\right)+4\kappa^2\tilde{\tau}^2\left(\kappa^4+6\eta^4\right)+12\left(\frac{\kappa'}{l'}\right)^2\left(8\eta^4-5\kappa^4\right)+\right.\\
        +\left.8\kappa\frac{\kappa''}{(l')^2}\left(\eta^4+4\kappa^4\right)\right]-\frac{16\sigma\eta}{l'\kappa^3}(8\tilde{\tau}\kappa'+\kappa\tilde{\tau}')+\frac{8\sigma^2}{\eta\kappa^2}\left(2\kappa^4+\kappa^2\tilde{\tau}^2+6\left(\frac{\kappa'}{l'}\right)^2+\kappa\frac{\kappa''}{(l')^2}\right)-\\
        -\frac{I_2^2}{ \eta \kappa^4} \left[2 \eta^4 + \kappa^4 (-1 + 2 \sigma^2)\right]-\frac{16\eta I_2}{\kappa^4} \left(- \eta^2 \tilde{\tau} + \sigma\kappa \frac{\kappa'}{l'}\right),
    \end{multline}
    \begin{multline}
        l'G_0\mathcal{D}_2=-\frac{16\eta\tilde{\tau}}{\kappa^2}\left(\frac{3\eta^4}{\kappa^4}+5\right)-\frac{32\sigma\kappa'}{l'\kappa\eta}\left(\frac{\eta^4}{\kappa^4}+1\right)-\frac{16\eta\tilde{\tau}\sigma^2}{\kappa^2}-\frac{64\kappa'\sigma^3}{l'\eta\kappa}+\\
        +\frac{8\eta I_2}{\kappa^6} \left(\eta^4 + \kappa^4 \sigma^2\right),
    \end{multline}
    \begin{equation}
        \mathcal{D}_3G_0(l')^2=-\frac{4}{\eta}\left(4+\frac{4\eta^8}{\kappa^8}+\frac{11\eta^4}{\kappa^4}\right)-\frac{4\sigma^2}{\eta}\left(7+\frac{4\eta^4}{\kappa^4}\right).
    \end{equation} 
\end{subequations}

The expressions for the quasisymmetric limit can be found in \citep{landreman2020} explicitly.

\section{Extra terms for Mercier stability in a QS stellarator} \label{sec:appLambda}
If one was to write the form of the Mercier criterion in an ideal quasisymmetric stellarator using the pressure gradient and the triangularity shaping on the up-down symmetric cross-section at the origin as free parameters, then we may write
\begin{equation}
    \frac{\epsilon^2\pi^2 D_\mathrm{Merc}}{|p_2|G_0^2}=\mathcal{T}_{|p|}|p_2|+\mathcal{T}_\delta \delta+\Lambda, \tag{\ref{eqn:DMercQSSplit}}
\end{equation}
where $\Lambda=\Lambda_2+\Lambda_0+\Lambda_{-2}$,
\begin{multline}
    \Lambda_2=\frac{1}{\eta^2\iota_0\kappa^6(3\eta^4 + 5\kappa^4) - 4\eta^4\kappa^8(I_2-\tau)}[\iota_0^3(\eta^4 + \kappa^4)(32\eta^8 + 11\eta^4\kappa^4-3\kappa^8) +\\
    +4\eta^2\kappa^6\tau(I_2-\tau)( 20\eta^4I_2-3(5\eta^4 + \kappa^4)\tau) -2\eta^2\iota_0^2\kappa^2(4(16\eta^8 + 8\eta^4\kappa^4 - 3\kappa^8)I_2- \\
    -(75\eta^8 + 50\eta^4\kappa^4 - 9\kappa^8)\tau) + \iota_0\kappa^4((201\eta^8 + 20\eta^4\kappa^4 + 3\kappa^8)\tau^2 + \\
     + 20\eta^4(-17\eta^4 + \kappa^4)\tau I_2 + 16\eta^4(8\eta^4 - 3\kappa^4)I_2^2)],
\end{multline}
\begin{multline}
    \Lambda_0=\frac{\eta^4\iota_0^2\kappa^3(-5\eta^4 + \kappa^4) + 4\eta^6\iota_0\kappa^5(\tau +3I_2)}{\eta^2\iota_0\kappa^3(\iota_0(3\eta^4 + 5\kappa^4) + 4\eta^2\kappa^2(\tau - I_2))} + \\
    +\frac{\iota_0(-11\eta^8 + 46\eta^4\kappa^4 - 3\kappa^8) + 4\eta^2\kappa^2(11\eta^4 - 3\kappa^4)(\tau - I_2)}{\eta^2\kappa^3(\iota_0(3\eta^4 + 5\kappa^4) + 4\eta^2\kappa^2(\tau - I_2))}\kappa'' - \\ 
    -\frac{32\eta^2[-\eta^2\iota_0 - \kappa^2(\tau - I_2)]}{\iota_0(\iota_0(3\eta^4 + 5\kappa^4) + 4\eta^2\kappa^2(\tau - I_2))}\tau'',
\end{multline}
\begin{equation}
    \Lambda_{-2}= \frac{8\eta^2\kappa''(\kappa^3 + 4\kappa'')}{\iota_0(\iota_0(3\eta^4 + 5\kappa^4) + 4\eta^2\kappa^2(\tau - I_2))},
\end{equation}
and the subscript denotes the scaling with rotational transform (or similar elements such as $\tau$ and $I_2$). We write these expressions using a normalised and scaled-out version of the equations, which eases notation. In this notation, to convert to their full explicit form, $\kappa$ and $\tau$ should be divided and $\eta$ multiplied by a factor of $\mathrm{d}l/\mathrm{d}\phi$. The current $I_2$ should be divided by $\mathrm{d}l/\mathrm{d}\phi$, as we assume $G_0=\mathrm{d}l/\mathrm{d}\phi$. The expression for $\Lambda$ should be then divided by $(\mathrm{d}l/\mathrm{d}\phi)^2$ to express it in a form consistent with the other terms in Eq.~(\ref{eqn:DMercQSSplit}). All quantities are evaluated at $\phi=0$, the location of the up-down symmetric cross-section of choice in our stellarator symmetric configuration. Note that whenever the denominators are small, the near-axis model will be very sensitive to change.
\par
If, instead of triangularity, we were to express the Shafranov shift explicitly, then we obtain,
\begin{equation}
    \mathcal{T}_\Delta=3\kappa\frac{(\alpha-1)+\bar{F}(1+\alpha)}{\alpha-\bar{F}(1+\alpha)},
\end{equation}
which reduces to the axisymmetric limit when $\bar{F}\rightarrow0$. Note that the tendency of the configuration will be for $\mathcal{T}_\Delta<0$, as this is the limit for a significant asymmetry $\bar{F}$. Unlike in the case of triangularity, this is the typical behaviour in a tokamak with vertical elongation. Bunching of surfaces on the inboard side (that is, $\Delta<0$) increases stability. Of course, there is a region in $(\alpha,~\bar{F})$ space in which $\mathcal{T}_\Delta>0$. Thus, the behaviour of a tokamak for a horizontally elongated cross-section may also be, in principle, achieved with the appropriate axis and parameter combination. 

\section{Validity of ideal quasisymmetric assumption} \label{sec:appQScheck}
To construct our analytic measure of stability, and understand the contribution of triangularity in quasisymmetric configurations, we took the simplifying assumption of ideal quasisymmetry. That is, we assumed $B_{20}$ to be constant. In practice, optimised quasisymmetric configurations such as those in Fig.~\ref{fig:shapeTriangPressMerc} are not ideal. That is to say, they all have a finite variation in $B_{20}(\phi)$. As presented in the main text, this means that there will be a mismatch between the estimate of stability from the ideal analysis and the full approach.
\par
\begin{figure}
    \centering
    \includegraphics[width=0.6\textwidth]{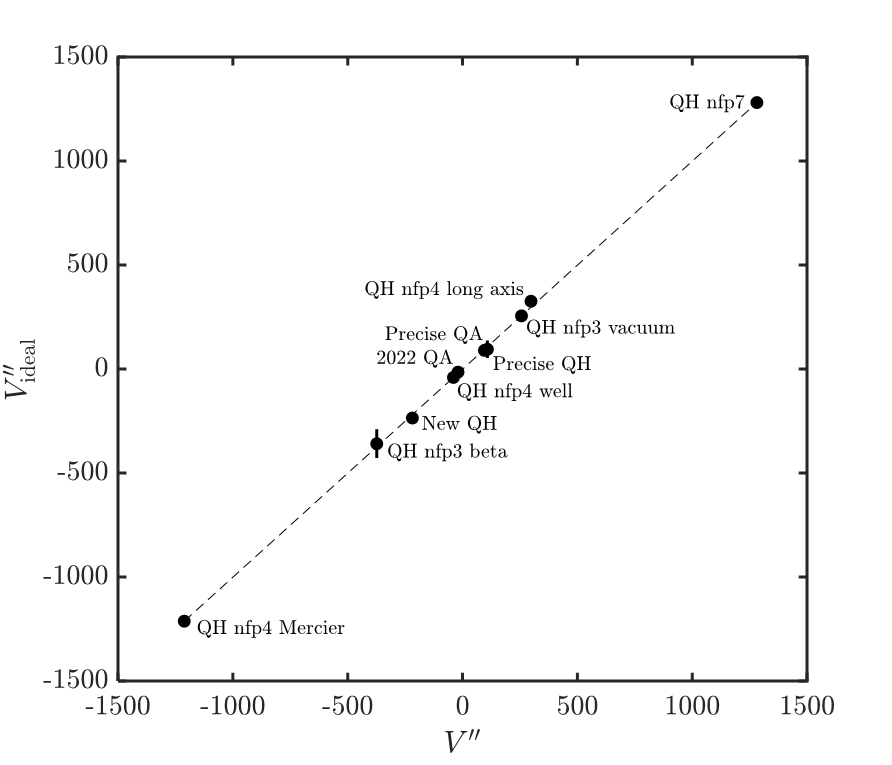}
    \caption{\textbf{Comparison of the magnetic well criterion between the full near-axis and idealised evaluation.} The plot compares the magnetic well criterion $V''$ for the configurations in Fig.~\ref{fig:shapeTriangPressMerc} assessed using the full near-axis description ($V''$) and the idealised quasisymmetric limit ($V''_\mathrm{ideal}$). The plot shows excellent agreement (within 30\% in all cases).} 
    \label{fig:checkVpp}
\end{figure}

To back the validity of Figure~\ref{fig:shapeTriangPressMerc} we should thus provide evidence of the ideal consideration being a fair descriptor of the stability. We present in Fig.~\ref{fig:checkVpp} a comparison between the magnetic well evaluated using the full near-axis form and the idealised scenario representing the stellarator by its up-down cross-section at $\phi=0$ (the cross-sections shown in Fig.~\ref{fig:shapeTriangPressMerc}, at $\phi=0$ as defined in the relevant papers). Only cases that showed agreement were kept, as only in those cases we expect the analysis to be trustworthy.  

\section*{Acknowledgements}
The author would like to acknowledge fruitful discussion with Carolin N\"{u}hrenberg, Wrick Sengupta, Per Helander, Gabe Plunk, Gareth R. Clark and  Ralf Mackenbach.


\section*{Declaration of interest}
The author reports no conflict of interest.


\bibliographystyle{jpp}

\bibliography{jpp-biblio.bib}

\end{document}